\documentclass[fleqn,usenatbib]{mnras}

\usepackage{newtxtext,newtxmath}
\usepackage{float}
\usepackage{tikz}
\usepackage{stfloats}
\usepackage{caption}

\usepackage[T1]{fontenc}
\usepackage{booktabs}
\DeclareRobustCommand{\VAN}[3]{#2}
\let\VANthebibliography\thebibliography
\def\thebibliography{\DeclareRobustCommand{\VAN}[3]{##3}\VANthebibliography}
\usepackage{xurl}

\usepackage{graphicx}	
\usepackage{amsmath}	
\usepackage{subcaption} 
\usepackage{tabularx}
\usepackage[dvipsnames]{xcolor}
\usepackage{siunitx} 

\title[SPIRIT on-sky performance]{On-sky performance and prospects of near-infrared CMOS detectors in exoplanet detection: first year of SPIRIT}

\author[C. Janó-Muñoz et al.]{C. Janó-Muñoz,$^{1}$\thanks{E-mail: cj467@cam.ac.uk}
M. J. Hooton,$^{1}$
P. P. Pedersen, $^{2}$
D. Queloz, $ ^{1,2}$
K. Barkaoui, $^{3, 4, 5}$
Y. T. Davis, $^{6}$
\newauthor
M. Gillon, $^{4}$
B. V. Rackham, $^{5}$
M. Timmermans, $^{6, 4}$
A. H. M. J. Triaud, $^{6}$
S. Zúñiga-Fernández $^{4}$
\\
$^{1}$Cavendish Laboratory, JJ Thomson Avenue, Cambridge CB3 0HE, UK\\
$^{2}$ETH Zurich, Department of Physics, Wolfgang-Pauli-Strasse 2, 8093 Zurich, Switzerland\\
$^{3}$ Instituto de Astrof\'isica de Canarias (IAC), Calle V\'ia L\'actea s/n, 38200, La Laguna, Tenerife, Spain \\
$^{4}$Astrobiology Research Unit, Université de Liège, 19C Allée du 6 Août, 4000 Liège, Belgium\\
$^{5}$Department of Earth, Atmospheric and Planetary Science, Massachusetts Institute of Technology, 77 Massachusetts Avenue, Cambridge, MA 02139, USA\\
$^{6}$School of Physics \& Astronomy, University of Birmingham,
Edgbaston, Birmingham B15 2TT, UK\\
}
\date{Accepted XXX. Received YYY; in original form ZZZ}

\pubyear{\the\year{}}

\begin{document}
\label{firstpage}
\pagerange{\pageref{firstpage}--\pageref{lastpage}}
\maketitle

\begin{abstract}

Ultracool dwarfs are promising targets for detecting and characterising temperate terrestrial exoplanets. Their flux peaks in the near-infrared, beyond the wavelength cut-off of the silicon-based CCDs used by ground-based surveys dedicated to observing them, such as SPECULOOS. We present the on-sky performance of SPIRIT (SPeculoos InfraRed Imager for Transits), an InGaAs CMOS detector on the Callisto telescope at the SPECULOOS Southern Observatory, during its first year of operation, from mid-2022 to 2023. We use 328 visits of 73 targets to optimise our CMOS-specific reduction process, including our approach to glow effects and bad-pixel handling. 
We find the measured precision to be consistent with our theoretical noise model, achieving sub-part-per-thousand precision under good observing conditions across targets spanning $J$-band magnitudes of 10--13. 
We further illustrate SPIRIT's capabilities with two case studies: a simultaneous SPIRIT $zYJ$ and CCD $I + z$ observation of a TRAPPIST-1\,h transit, and monitoring observations of the fast-rotating dwarf LP~944-20. Our results confirm that InGaAs infrared CMOS detectors are a cost-effective and competitive technology for ground-based transit photometry of cool stars, and that they show promising prospects to become a core part of the next generation of instruments.

\end{abstract}




\section{Introduction}

Ultracool dwarfs (UCDs) -- defined as objects with spectral types M7 and later, and cooler than 2700\,K -- are advantageous targets when studying habitable-zone exoplanets. The proximity of their habitable zones and their small radii translate to more frequent and deeper transits, with higher signal-to-noise ratios, compensating for current observational biases. This then makes them optimal targets for atmospheric measurements, particularly relevant in the current era of JWST \citep[][]{Gardner2023} and the upcoming era of ELT's ANDES spectrograph \citep[][]{Palle2025}. 

One such star is TRAPPIST-1, host to seven rocky transiting exoplanets \citep[][]{Gillon2017}, and currently the most observed exoplanet host with JWST\footnote{TrExoLiSTS; \url{https://www.stsci.edu/~nnikolov/TrExoLiSTS/JWST/trexolists.html}}. This planetary system was discovered with TRAPPIST (TRAnsiting Planets and PlanetesImals Small Telescope; \citealt[][]{Gillon2011}), a prototype programme dedicated to the detection of rocky habitable-zone exoplanets around UCDs. The success of this prototype led to the commissioning of SPECULOOS (Search for habitable Planets EClipsing Ultra-cOOl Stars; \citealt[][]{Burdanov2018, Delrez2018}). 

SPECULOOS comprises the SPECULOOS South and North Observatories (SSO and SNO) with four and one telescopes respectively, and the Saint-Ex telescope \citep[][]{SaintEx}. All six are 1-m robotic equatorial telescopes originally equipped with Andor-iKon 2\textit{k} $\times$ 2\textit{k} charge-coupled device (CCD) cameras with a pixel scale of 0.35\arcsec, yielding a total field of view of $12\arcmin \times 12\arcmin$. The observations are focused on the red end of the visible and the start of the near-infrared (NIR) regime, to approach the wavelengths where UCDs emit most strongly, with a custom $I + z$ filter for observations beyond $700$\,nm (see \citealt[][]{Delrez2018, Burdanov2018}). The flux of UCDs, however, peaks in the NIR at wavelengths longer than those the SPECULOOS CCDs can detect (e.g., the emission of TRAPPIST-1, an M8 dwarf, peaks around 1.2\,$\text{\textmu}$m; \citealt[][]{Davoudi2024}). The effects of stellar limb darkening, which are degenerate with planet radius, are also less significant in the NIR, allowing a more precise measurement of transit parameters when observing in this regime.

Moving to these wavelengths presents significant challenges in ground-based astronomy. First, silicon-based detectors, such as CCDs, have an inherent wavelength cut-off at 1.1\,$\text{\textmu}$m due to the silicon band gap at 1.12\,eV. Second, on-sky, the NIR band transmission is limited by telluric water absorption bands, especially around 0.9, 1.1 and 1.4\,$\text{\textmu}$m (see \citealt[][]{Pedersen2023}). 

The current generation of NIR instruments usually solves the silicon cut-off issue by using HgCdTe or InSb detectors. The use of these materials requires cryogenic cooling, increasing the installation and maintenance costs of the instrument. A ground-based survey similar to SPECULOOS is the Perkins INfrared Exosatellite Survey \citep[PINES;][]{Tamburo2022}, which searches for exoplanets around L and T dwarfs with an InSb array detector operating at 33.5\,K \citep[][]{Clemens2007}. A different approach to NIR detectors is the use of fibres for joint photometry and spectroscopy, as employed by ExTrA \citep[Exoplanets in Transits and their Atmospheres;][]{Bonfils2015}. 

A compromise between performance and cost can be reached with InGaAs complementary metal-oxide-semiconductor (CMOS) detectors, which do not require cryogenic cooling to operate. Standard InGaAs has a wavelength cut-off at 1.7\,$\text{\textmu}$m, and so it is optimised for observations at the short-wavelength end of the NIR regime. This means that, while not accessing the 2--2.5\,$\text{\textmu}$m range of the NIR, it is insensitive to long-wavelength thermal radiation such as its own self-emission, which would otherwise be a significant source of dark current. The temperature required to operate with InGaAs can be reached through non-cryogenic (thermoelectric) cooling, which lowers both installation and maintenance costs. For a comprehensive comparison of operating temperatures of different NIR detector materials, see \citet{Rogalski2002}.

CMOS detectors \citep[][]{CMOS} differ from CCDs \citep[][]{CCD} in structural design and operation. CCD detectors work by transferring pixel charges across the sensor array to a readout node, where they are then converted to voltage and processed. This sequential transfer offers high image quality with low noise levels but a slow readout. In contrast, for CMOS sensors, each pixel is equipped with its own amplifier. This allows parallel readout across the entire array, enabling significantly faster readout speeds. The fast readout provides an advantage for high-cadence time-series observations. See \citet{Hansen2008} and \citet{Simcoe2019} for a detailed view of InGaAs CMOS detectors and their application in photometry. Recent uses of CMOS sensors in astronomy show these instruments can achieve photometric precision and noise levels comparable to those of their CCD counterparts \citep[][]{Alarcon_2023}. A recent relevant example, but in the visible wavelength regime, can be seen in NGTS's Marana camera \citep[][]{Apergis2026}.

In this work, we study the performance of the InGaAs CMOS detector SPIRIT (SPeculoos InfraRed Imager for Transits). This instrument was first introduced by \citet{SPIRIT} and has been used in planet-validation studies \citep[e.g.][]{Triaud2023, JanoMunoz2025, Dransfield2025}. SPIRIT is a 1280 $\times$ 1024 pixel near-infrared InGaAs CMOS detector installed on the Callisto telescope in the SSO. We attempt to mitigate the other main challenge of ground-based NIR photometry -- atmospheric water absorption -- by both instrument design and data calibration. SPIRIT observes in the 0.81--1.33\,$\text{\textmu}$m range with a custom wide-pass filter, $zYJ$, that minimises the effects of precipitable water vapour (PWV), as well as narrower $J$-band and $H$-band filters for the coldest targets. We simultaneously record the atmospheric PWV in Paranal during observations and correct for its effect in post-processing. 

SPIRIT initially operated from 2022 to 2023, cooled to \SI{-60}{\degreeCelsius}. It was then uninstalled due to a cooling failure, returned on-sky in 2025, and has operated at \SI{-50}{\degreeCelsius} since then. We discuss the effect of this temperature change in more detail in Section \ref{subsec:temp}. The initial performance analysis by \citet{SPIRIT} showed that noise models aligned well with on-sky observations, and demonstrated superior performance compared to the CCD instruments for targets cooler than 2550\,K. In this work, we extend the analysis to the first year of SPIRIT observations, and discuss our strategies for further optimisation. 

We describe our target sample in Section \ref{sec:inst}. In Sections \ref{sec:methods} and \ref{sec:performance}, we explain our calibration methods, optimisation, and global performance results. We then show two case studies, the observations of a TRAPPIST-1\,h transit and the fast-rotating M9.5 dwarf LP~944-20 in Section \ref{sec:target}. Finally, we discuss future prospects in Section \ref{sec:prospects} and present our conclusions in Section \ref{sec:conclusion}.

\section{Target list}
\label{sec:inst}

\begin{figure*}
    \centering
    \includegraphics[width=\linewidth]{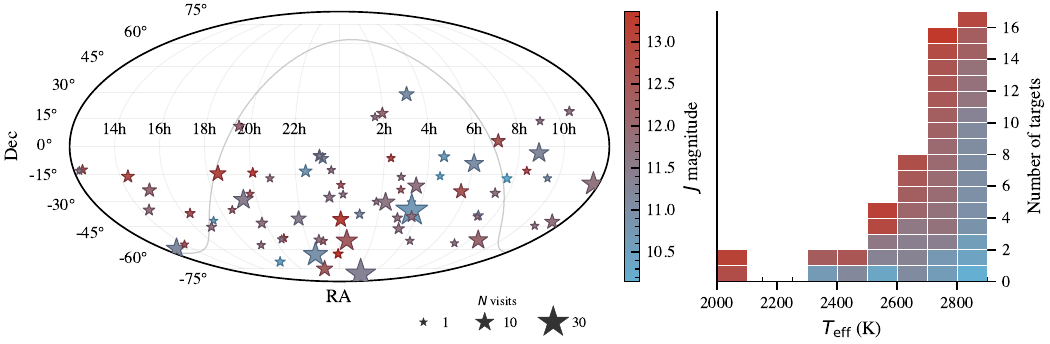}
    \caption{Sample of targets used in the analysis in this work, observed in 2022 and 2023 with SPIRIT, shown as a map of their position in the sky (left) and as a histogram (right). The size of the star icon scales with the number of observations taken of that target. The colour maps the target $J$ magnitude, and the faint line in the sky map represents the Galactic plane. The histogram counts the number of targets (excluding TOIs) at different effective temperatures in 100-K bins.}
    \label{fig:skymap}
\end{figure*}

During 2022 and 2023, 476 observations were taken. Out of these, 328 (69\%) were successfully reduced and are used in this work. We found that the main cause behind unsuccessful reduction was an insufficient number of stars in the field for successful astrometry and differential photometry. Our method to counter this is described in Section \ref{subsec:goibniu}. These observations contain a sample of 73 targets, which were selected both to test the capabilities of SPIRIT for the coolest targets in the SPECULOOS target list and to be representative of the overall survey target list to compare performance across instruments.
Thirty-five observations (11\%) of 20 different targets (27\%) were taken simultaneously with SPIRIT and at least one SSO CCD camera to compare performance under the same observing conditions; a comparison between SPIRIT and the SSO CCDs is also presented in \citet{SPIRIT}, and we examine one such observation in detail in Section \ref{sec:target}. We also note that 16 of these targets are follow-ups of TESS Objects of Interest (TOIs; \citealt[][]{TOI}), since validation and characterisation of TESS targets is part of the SPECULOOS programme \citep[][]{Sebastian2021}. We used the complete sample of 73 targets for optimising data reduction, but excluded TOIs from the analysis in Section \ref{subsec:globalanalysis} to focus on the main-programme targets for which SPIRIT is optimised.

The targets have an average $J$-band magnitude of 11.8 and an average effective temperature of 2716\,K.
The coolest single object is the L2 dwarf Sp0828-1309 at 2088\,K; Sp0004-4044 appears cooler in blended light (2041\,K) but is an unresolved pair of L4.5 dwarfs, each with $T_\mathrm{eff} \approx 1725$\,K (see Figure~\ref{fig:radii}). 
A sky map of our target list is shown in Figure \ref{fig:skymap}, and the full list can be consulted in Appendix \ref{appendix}.

\section{CMOS Optimisation}
\label{sec:methods}

SPECULOOS data reduction is performed by the SPECULOOS pipeline \citep[][]{Murray2020}, which uses standard photometric reduction techniques. The calibration steps, however, differ when moving from the visible to the NIR and from CCDs to a CMOS. For this work, the data were reduced and calibrated using a minimal local version of the SPECULOOS pipeline built with the \texttt{prose}\footnote{\texttt{prose}: \url{https://github.com/lgrcia/prose}} software \citep[][]{prose_2022}, and the optimisation steps have since been incorporated into the pipeline for all subsequent observations.

\subsection{Dark current and thermal glow}

An effect particularly prominent in CMOS detectors is thermal glow (sometimes referred to as amplifier glow). Each pixel's sensor and amplifier, as well as the detector's peripheral circuitry, emits its own radiation, which is absorbed by neighbouring pixels and read as signal, creating a glow-like effect. This effect is driven primarily by the electronic activity during readout rather than by thermal generation, meaning it would persist even if the detector were cooled down further \citep[][]{Tam1984, Regan2020}. Since it has a much weaker temperature dependence than dark current, we treat them as separate components. Approaches to mitigate this effect include extracting the glow pattern with long-exposure frames \citep[][]{Marana2025} or identifying it and removing it through statistical techniques \citep[][]{Zhang2023}. 

Given the specifications and use of SPIRIT, we opt for a different approach. Dark frames are taken at the same exposure time as the science frames. These matched-exposure dark images are median-combined to construct the master dark frame, which contains both the true dark current and the exact glow pattern that will be present in the science frames. This is then subtracted from the science frames in the calibration process. This removes the glow as long as dark and science frames are read with the same readout mode and at the same operating temperature, as is the case with SPIRIT. 

The true dark current is measured in SPIRIT following the same process as with the SPECULOOS CCDs. Dark images are taken at increasing exposure times to measure the increasing response. Although not used directly in the calibration, the true dark current is used for noise characterisation and to detect bad pixels (Section \ref{subsec:badpixels}).

\subsection{Bias and flat correction}
\label{subsec:bias}

In CCD photometry, bias frames with zero-second exposures are taken to calibrate for the detector's inherent electronic offset. SPIRIT, however, uses the correlated double sampling (CDS) readout method: the signal from each amplifier is read at the beginning and at the end of the exposure, and the difference is given as the bias-free collected flux \citep[see][]{Janesick2001}. The CDS readout also makes zero-second exposures impossible with SPIRIT, so purely bias frames free of dark current cannot be obtained. The readout electronics of a CMOS, however, cause characteristic row-by-row and/or column-by-column patterns in the images, which are commonly corrected via a ``pseudo-bias'' calibration \citep[see e.g.][]{Layden2025}. 

As mentioned in the previous subsection, we use matched-exposure dark frames to correct for the true dark current and the glow. The master dark frames already contain the row and column artefacts of the detector, so it is sufficient to use them alone. This can only work, however, if our flat-field correction follows the same methodology. Flat-field correction is performed in SPIRIT the same way as in our CCDs; we use the sky at dusk and dawn as a uniform light source \citep[][]{Murray2020}. To be consistent in removing the glow and not requiring bias correction, flat frames also have to be corrected with dark frames matching their exposure time. For the 2022--2023 data analysed here, dark frames matching the flat exposure times were not available, so we instead corrected the flats for the row-by-row pattern using short bias-like exposures. This is the only step in which the reduction presented in this work deviates from the procedure described above, and it affects the flat fields alone.

\subsection{Bad pixels}
\label{subsec:badpixels}
\begin{figure}
    \centering
    \includegraphics[width=0.99\linewidth]{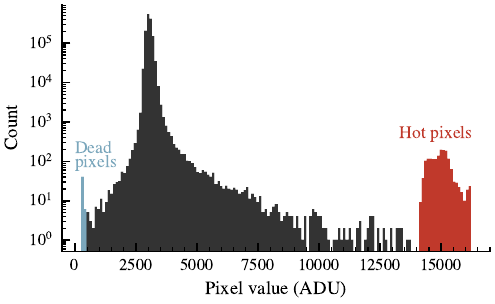}
    \caption{Histogram of the pixel values in analogue-to-digital units (ADU) in the median of nine 10\,s dark frames, taken on 22 December 2022. The peak at the lower values in blue (labelled as dead pixels), and the red population at the high end (labelled hot pixels) show two types of bad pixels found in the detector that are easily identified in the dark and flat frames. Hot pixels are selected at the detector's saturation threshold of 13,820 ADU.}
    \label{fig:pixelhist}
\end{figure}

In the first analysis of SPIRIT \citep[][]{SPIRIT}, we found around 4,000 bad pixels in the detector, corresponding to 0.3\% of all pixels. We have identified the types of bad pixels in SPIRIT and the effect they have on photometry. The largest population of bad pixels consists of hot/saturated pixels, which we detect in the dark frames. A smaller number of dead pixels -- those with no or too-low response -- are identified during the flat-field correction stage. The distribution of pixel values showing these two populations of bad pixels is displayed in Figure \ref{fig:pixelhist} for a set of dark frames. By measuring the response of each pixel as a function of exposure time, we can further identify pixels that are stuck at an ADU value or show either a negative or a random evolution with exposure time.

\begin{figure}
    \centering
    \includegraphics[width=0.95\linewidth]{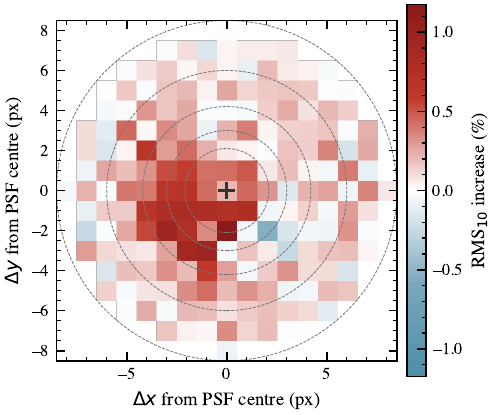}
    \caption{Effect of bad pixels as a heat map on the SPIRIT PSF for the TRAPPIST-1 observation on 28 August 2022. The colour of each pixel indicates the change in 10-minute binned RMS in the final light curve if that pixel is flagged as bad and interpolated over. The cross in the centre indicates the centroid and the dashed circles represent aperture sizes.}
    \label{fig:badmap}
\end{figure}

To quantify how bad pixels can affect the photometry, we considered the case of a single bad pixel falling within the point-spread function (PSF) of our target. We used an observation of TRAPPIST-1 from 28 August 2022, in which there were no bad pixels inside the PSF of either TRAPPIST-1 or the comparison stars. Our first step was to perform our standard data reduction and calculate the 10-minute binned RMS, in this case 0.5\,ppt. 
We then repeated this process by setting one pixel inside the PSF to zero, to be flagged during calibration. This artificial bad pixel was corrected in the reduction through masking and interpolation \citep[][]{Popowicz2013}, and we obtained a new 10-minute binned RMS. Repeating this systematically for each of the pixels in the PSF, we measured the change in precision caused by bad pixels at each of the positions. The result, shown in Figure \ref{fig:badmap}, is a small change in precision for bad pixels located only within the wider apertures, but a 1.2\% worsening for pixels near the centre and where the PSF gradient is steepest. We note that the apparent improvement in precision in some pixels suggests that bad pixel correction can cause an underestimation of the uncertainties. This is nevertheless a best-case scenario, with only one bad pixel in the PSF of the target, none in the comparison stars, and good tracking. In the cases of multiple bad pixels (particularly neighbouring ones) present in the PSFs, or bad tracking causing them to shift in and out of the apertures, we expect a larger effect on the photometry. A quantitative analysis of this is beyond the scope of this work but will be explored in the future. We therefore conclude that the best approach is to avoid bad pixels altogether, and describe our method to do this in Section \ref{subsec:goibniu}.

\section{On-sky performance}
\label{sec:performance}

\subsection{Global analysis}
\label{subsec:globalanalysis}

We systematically reduced and analysed the data from the 2022 and 2023 observations with our custom-made pipeline. For each observation, we measured the 10-minute-binned root-mean-square (RMS) scatter in parts per thousand (ppt), and compared it with a theoretical noise prediction given by 
\small
\begin{equation}
    \sigma_{\text{theory}} = \frac{\sqrt{N_\star + \sigma_{\text{scint}}^2 + n_{\text{pixel}} \times (N_{\text{sky}} + N_{\text{thermal}} + N_{\text{dark}} + N_{\text{read}}^2)}}{N_\star}
\end{equation}
\normalsize
where $N_\star$, $N_{\text{sky}}$, $N_{\text{thermal}}$ and $N_{\text{dark}}$ are the photoelectron counts from each source, $N_{\text{read}}$ is the read noise per pixel in electrons, $n_{\text{pixel}}$ is the number of pixels inside the photometric aperture, and $\sigma_{\text{scint}}$ is the noise contribution from scintillation, caused by atmospheric turbulence \citep[][]{SPIRIT}. 
The stellar and sky contributions ($N_\star$ and $N_{\text{sky}}$) were obtained using the \texttt{mphot} package\footnote{\texttt{mphot}: \url{https://github.com/ppp-one/mphot}}, which calculates the expected stellar flux and sky radiance as a function of airmass and PWV, based on ESO Sky Models \citep[][]{ESOsky}. 

\begin{figure}
    \centering
    \includegraphics[width=0.99\linewidth]{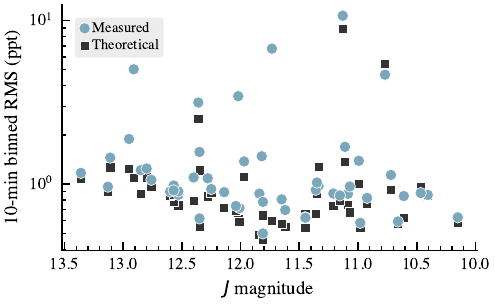}
    \caption{Mean RMS of each of the 52 main programme targets in the SPIRIT 2022--2023 observing run as a function of $J$-band magnitude. The blue (circular) points correspond to the measured 10-minute-binned RMS, while the black (square) points are the theoretically expected precision values of each target.}
    \label{fig:rms}
\end{figure}

During the differential photometry stage, the target noise is then combined with the noise from the comparison stars, which is a significant contribution in SPIRIT due to the limited number of available stars in the field. The total precision is calculated in 10-minute bins to limit the contribution of long-term variability throughout the observing window. 

\begin{figure}
    \centering
    \includegraphics[width=0.95\linewidth]{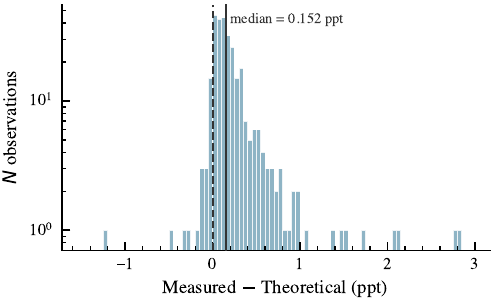}
    \caption{Histogram showing the difference between measured and predicted 10-minute-binned RMS for every observation with SPIRIT used in this work. The dashed line represents the zero point, corresponding to full agreement, while the solid line is placed at the median difference.}
    \label{fig:rmshist}
\end{figure}

The resulting binned RMS as a function of $J$-band magnitude is shown in Figure \ref{fig:rms}, where we compare it with our theoretical prediction. The difference between our measured precision and our noise model is shown as a histogram in Figure \ref{fig:rmshist}. The measured RMS exceeds the prediction by a median of 0.15\,ppt, with a wide distribution. Outliers at the high end correspond to targets observed under non-optimal observing conditions, while those showing better precision than predicted are likely due to the limitations of our stellar noise models from \texttt{mphot}.

\begin{figure}
    \centering
    \includegraphics[width=0.99\linewidth]{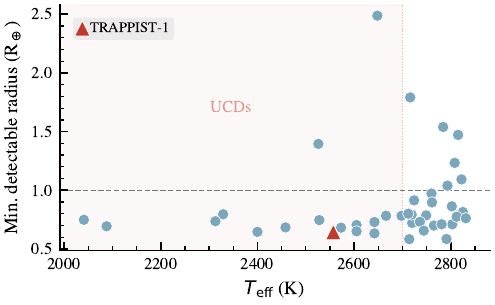}
    \caption{Minimum detectable radius, in Earth radii, around each SPIRIT 2022--2023 target as a function of stellar effective temperature, calculated from the mean 10-minute-binned RMS of each target for a transit depth of four times the RMS. TRAPPIST-1 is highlighted in red (triangle), and the UCDs in our sample are enclosed in the red (darkened) region. The dashed line marks 1 R$_\oplus$. The point at 2041\,K corresponds to Sp0004-4044 (GJ 1001 B and C; \citealt[][]{GJ1001}), an unresolved pair of L4.5 brown dwarfs in a triple system with an M4V primary (GJ 1001 A). The temperature here reflects the blended light of the two sources, each with $T_{\text{eff}} \approx 1725$\,K \citep[][]{GJ1001temp}.} 
    \label{fig:radii}
\end{figure}

Based on the measured precision, we can empirically predict the smallest detectable planetary radius for each star in our sample of 52 main programme targets, shown in Figure \ref{fig:radii}. We define this as the planet radius producing a transit depth of four times the 10-minute-binned RMS, $R_{\text{min}} = R_\star \sqrt{4\,\sigma_{10}}$. We estimate that we could detect the transits of Earth-sized planets around most of our targets. We draw attention, however, to the high activity of UCDs, particularly the coolest targets, which poses additional challenges for transit detection.

\subsection{Long-term performance}
\label{subsec:temp}

Our global analysis and optimisation process up until this point has been based on the instrument's performance when cooled to \SI{-60}{\degreeCelsius}. As mentioned previously, since 2025, the instrument has been observing at \SI{-50}{\degreeCelsius} to preserve the detector's longevity. We studied the effect this had on the precision. Increasing the detector temperature increases the dark current, as described by the Arrhenius equation. When applied to a change of temperature in a CCD, this equation takes the form of the ratio
\begin{equation}
    \alpha = \frac{N_{\rm dark}(T_2)}{N_{\rm dark}(T_1)} = \left(\frac{T_2}{T_1}\right)^2 \exp{\left[-\frac{E_g}{2k_B}\left(\frac{1}{T_2} - \frac{1}{T_1}\right)\right]},
\end{equation}
where $T_1$ and $T_2$ are the temperatures of the detector before and after, and $E_g$ is the band-gap energy \citep[][]{Janesick2001, Widenhorn2022}. A CMOS detector follows the same physics: thermal energy excites the electrons from the valence to the conduction band, separated by the band-gap energy of the photosensitive material (InGaAs instead of silicon in this case). Adopting $E_g \approx 0.73$\,eV for InGaAs, corresponding to its 1.7\,$\text{\textmu}$m cut-off, equation 2 gives $\alpha \approx 2.7$ for the change from \SI{-60}{\degreeCelsius} to \SI{-50}{\degreeCelsius}, consistent with the empirical doubling of InGaAs dark current for every \SI{7}{\degreeCelsius} increase \citep[][]{Sullivan2014}. This increase will apply only to thermally generated dark current, and will not affect sources with a much weaker temperature dependence, such as glow. This means the total increase in $N_{\rm dark}$ will be less than the factor of 2.7 we have calculated. We also expect the contribution to be modest since, as detailed in \citet[][]{SPIRIT}, SPIRIT is read-noise limited, not dark-current limited.

By increasing the value of $N_{\rm dark}$ in our noise estimation described in the previous section, and keeping the excess noise from Figure \ref{fig:rms}, we can estimate the effect this has on our precision. We initially use the 2.7 factor, to be as conservative as possible in our calculation. We estimate a median increase of 1\% in the 10-minute RMS of our observations (comparable to the effect of a single bad pixel in the target PSF, Section \ref{subsec:badpixels}), but reaching up to a 68\% increase in the case of the faintest targets in our sample.  

We verified this estimate by measuring the dark current noise contribution in the data taken in 2025 and 2026. Here, the dark current increased by a factor of 1.4--1.7 compared to the 2022 and 2023 observations, indicating that 24--41\% of the dark current is thermally generated. Since this is smaller than the conservative factor of 2.7 assumed above, the estimated increases in RMS are upper limits. We can therefore apply the conclusions of our global analysis to the data taken after 2025, with the caveat that the faintest targets are the most affected.

\section{Target-specific performance}
\label{sec:target}

\subsection{Case study I: TRAPPIST-1\,h}
\label{subsec:trappist1h}
Our first case study to test the scientific capabilities of SPIRIT is an observation of TRAPPIST-1\,h taken simultaneously by SPIRIT and the Andor CCD camera on the SSO's Europa telescope. 
We expect TRAPPIST-1, with a temperature of $\sim$2560\,K \citep[][]{Gillon2017}, to be at the boundary of where SPIRIT outperforms the SPECULOOS CCDs, since for cooler targets SPIRIT collects more photons per unit time, and it observes at a higher cadence \citep[][]{SPIRIT}.  

\begin{figure*}
\centering
\begin{minipage}[t]{0.49\textwidth}
    \vspace{0pt}
    \centering
    \includegraphics[width=0.99\linewidth]{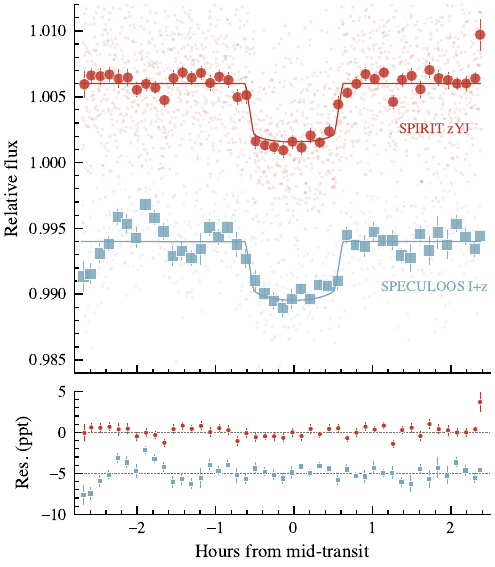}
    \captionof{figure}{Simultaneous observation of TRAPPIST-1\,h from 28 August 2022. The top panel shows the SPIRIT data in the $zYJ$ band (red circles) and the Andor CCD data in the $I+z$ band obtained with Europa (blue squares). The solid lines are the fitted transit models using \texttt{batman} \citep[][]{batman}, after detrending both datasets against environmental parameters. The bottom panel shows the residuals of both datasets in ppt.}
    \label{fig:trappist1h}
\end{minipage}
\hfill
\begin{minipage}[t]{0.49\textwidth}
    \vspace{0pt}
    \centering
    \captionof{table}{Summary of the data quality of the joint TRAPPIST-1\,h observation of 28 August 2022, taken simultaneously with SPIRIT ($zYJ$) and the Europa CCD ($I+z$). Noise values are in ppt. The expected white-noise scatter is $\sigma_{\text{exp}} = \sigma_1 N^{-1/2} [M/(M-1)]^{1/2}$, for $N$ points per bin and $M$ bins \citep[][]{Winn2008}. Uncertainties were propagated analytically. For $zYJ$, the result is consistent with no red noise and an upper limit is quoted for its amplitude. The $\beta$ factor was measured on the 10-minute bins.\label{tab:trappist1h_data_quality}}
    \begin{tabularx}{\linewidth}{l>{\centering}X>{\centering\arraybackslash}X}
    \toprule
        Filter & $zYJ$ & $I+z$ \\ \midrule
        $N_{\rm points}$                                & 2001 & 604  \\
        Baseline (hours)                                & 5.1  & 5.1  \\
        Cadence (s)                                     & 9.0  & 30.1 \\ \midrule
        Unbinned RMS $\sigma_1$                         & $3.83\pm0.07$ & $2.99\pm0.10$ \\
        10-min binned RMS $\sigma_{\text{N}}$           & $0.53\pm0.08$ & $0.95\pm0.14$ \\
        Expected binned RMS $\sigma_{\text{exp}}$       & 0.47 & 0.68 \\
        $\beta$ factor                                  & $1.14\pm0.18$ & $1.54\pm0.24$ \\
        Red noise $\sigma_{\text{red}}$                 & $<0.5$ & $0.72\pm0.19$ \\
    \bottomrule
    \end{tabularx}

    \vspace{2mm}
    \includegraphics[width=0.968\linewidth]{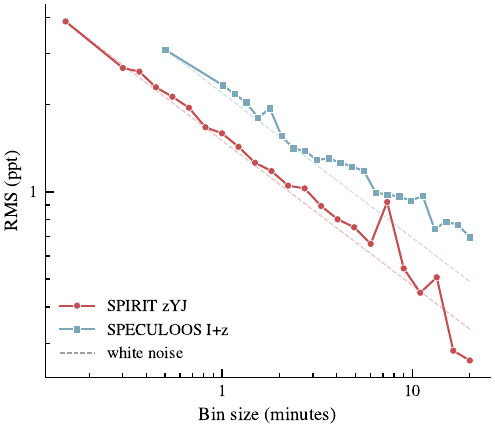}
    \captionof{figure}{Red-noise plot showing the RMS of the TRAPPIST-1\,h simultaneous SPIRIT (red, circle) and CCD (blue, square) observation at different bin sizes. The dashed lines show the expected white-noise scaling of each dataset. The aperture radii used were 6.0 and 11.3\,pixels for SPIRIT and $I+z$ respectively.} 
    \label{fig:rednoise}
\end{minipage}
\end{figure*}

We observed the target simultaneously on 28 August 2022. Given that the observations were taken at the same time, and from the same location, we expect differences in precision and data quality to arise mainly from differences between the two instruments (including their telescopes and comparison-star ensembles) and from the different sensitivity of each band to PWV. Some atmospheric effects, such as PWV and airmass, are wavelength-dependent, which we take into account when detrending our datasets. The data gathered are shown in Figure \ref{fig:trappist1h}.

As summarised in Table \ref{tab:trappist1h_data_quality}, the 10-minute binned precision on this night is better with SPIRIT than with the CCD ($0.53$ versus $0.95$\,ppt). The unbinned RMS is higher in $zYJ$ ($3.83$ versus $2.99$\,ppt), but at a cadence of 9.0\,s rather than 30.1\,s; scaled to the same integration time, the SPIRIT scatter corresponds to 2.09\,ppt, lower than that of the CCD. The higher cadence therefore gives SPIRIT the edge once the data are binned. The higher sampling rate also allows for more effective removal of environmental effects in the detrending process.

We measured the contribution of red noise to the precision for both wavelengths. White noise is uncorrelated in time, and thus averaging a number of data points $N$ into a bin reduces it as $1 / \sqrt{N}$. Since SPIRIT's $zYJ$ filter is designed to minimise the effect of PWV, we expect it to follow white noise more closely than $I + z$. We display this in Figure \ref{fig:rednoise} as a red-noise plot to visualise and measure how it deviates from white noise \citep[see][]{Pont2006, Winn2008}, and quantify it with the factor $\beta = \sigma_{\text{N}}/\sigma_{\text{exp}}$ \citep[][]{Winn2008}. The SPIRIT data are consistent with white noise, with $\beta = 1.14 \pm 0.18$ and an upper limit of $\sigma_{\text{red}} < 0.5$\,ppt. The $I+z$ data, on the other hand, deviate progressively from white noise with increasing bin size, with $\beta = 1.54 \pm 0.24$ and a red-noise contribution of $\sigma_{\text{red}} = 0.72 \pm 0.19$\,ppt. Of the factor of 1.8 between the binned precisions of the two instruments, a factor of 1.45 is expected from white noise alone ($\sigma_{\text{exp}}$), with the remainder due to the red noise in $I+z$. This is consistent with the result of \citet[][]{SPIRIT} that the PWV-optimised filter minimises the red noise.

\subsection{Case study II: LP~944-20}
\label{subsec:BD}
LP~944-20 is a fast-rotating M9.5/L0 brown dwarf \citep[][]{Tinney1998, McCaughrean2002}, with a rotation period of 3.8\,h measured using TESS \citep[][]{Fetherolf2023, Lambier2025}. This target was visited by SPIRIT on 31 nights during the 2022--2023 set of observations, specifically from October to December of 2022. 
We measured its rotation period in the NIR to compare it with the TESS results and test the activity-monitoring capabilities of SPIRIT.

From the 31 visits, we removed those interrupted by weather that resulted in observing sequences shorter than 1 hour, leaving 25 nights. We used a simple method to recover the rotation signal from the photometric variation, first computing a Lomb--Scargle periodogram \citep[][]{Lomb, Scargle} to find a first estimate of the period, adopting the Rayleigh width ($P^2/T$, for a baseline $T = 46$\,d) as the uncertainty. We then applied a sine-plus-cosine fit at
the estimated period, solved simultaneously with detrending in a single weighted-least-squares design matrix to minimise absorption of the signal by the baseline. The detrending baseline model was constructed independently for each night as a linear combination of the environmental parameters airmass, sky background, full width at half-maximum (FWHM) of the PSF, and displacement on the $X$ and $Y$ axes of the detector. For each light curve, we used the Bayesian Information Criterion \citep[BIC;][]{BIC} to decide which parameters to use in the baseline. We always included in the baseline an offset as a free parameter, to account for the change in observing conditions from night to night. 

We initially found a rotation period $P_{\rm rot} = 3.8415 \pm 0.0133\,\text{h}$, consistent with the value of $P_{\rm rot} = 3.8446^{\scriptscriptstyle +0.0065}_{\scriptscriptstyle -0.0096}\,\text{h}$ found by \citet[][]{Lambier2025}. We refined our $P_{\rm rot}$ measurement further by scanning a grid of possible $P_{\rm rot}$ values to minimise the phase difference between different nights (method adapted from \citealt[][]{Stellingwerf1978} and \citealt[][]{Harding2013}). This is particularly useful for our data since the gaps between observations correspond to multiple rotations. We found a final rotation period $P_{\rm rot} = 3.8411 \pm 0.0044\,\text{h}$, a factor of three more precise than our initial estimate and consistent with \citet[][]{Lambier2025} to within $0.4\sigma$.
The phase-folded light curve in Figure \ref{fig:phasefold} shows the variability recovered from the SPIRIT data; the per-night fits are shown in Appendix \ref{appendixb}. This case demonstrates the feasibility of characterising fast-rotating UCDs with SPIRIT, and of using it in the future to disentangle stellar variation from planetary signals with multi-wavelength observations. 

\begin{figure}
    \centering
    \includegraphics[width=0.99\linewidth]{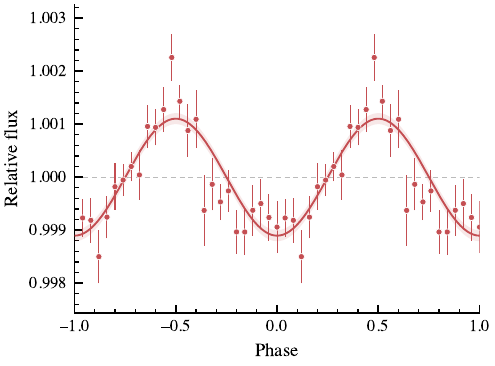}
    \caption{Phase-folded observations of LP~944-20 with SPIRIT, including 25 visits over the span of October to November 2022. The data are folded on the final rotation period, shown over two cycles for easier visualisation, and binned to 7 minutes (red circles). The solid line shows the best-fit rotation model, and the shaded band shows the 1$\sigma$ uncertainty on the amplitude.}
    \label{fig:phasefold}
\end{figure}
\section{Prospects}
\label{sec:prospects}

\subsection{Further optimisation}
\label{subsec:goibniu}

\begin{figure*}
\centering
\begin{subfigure}[t]{.48\textwidth}
    \centering
    \includegraphics[height=7cm]{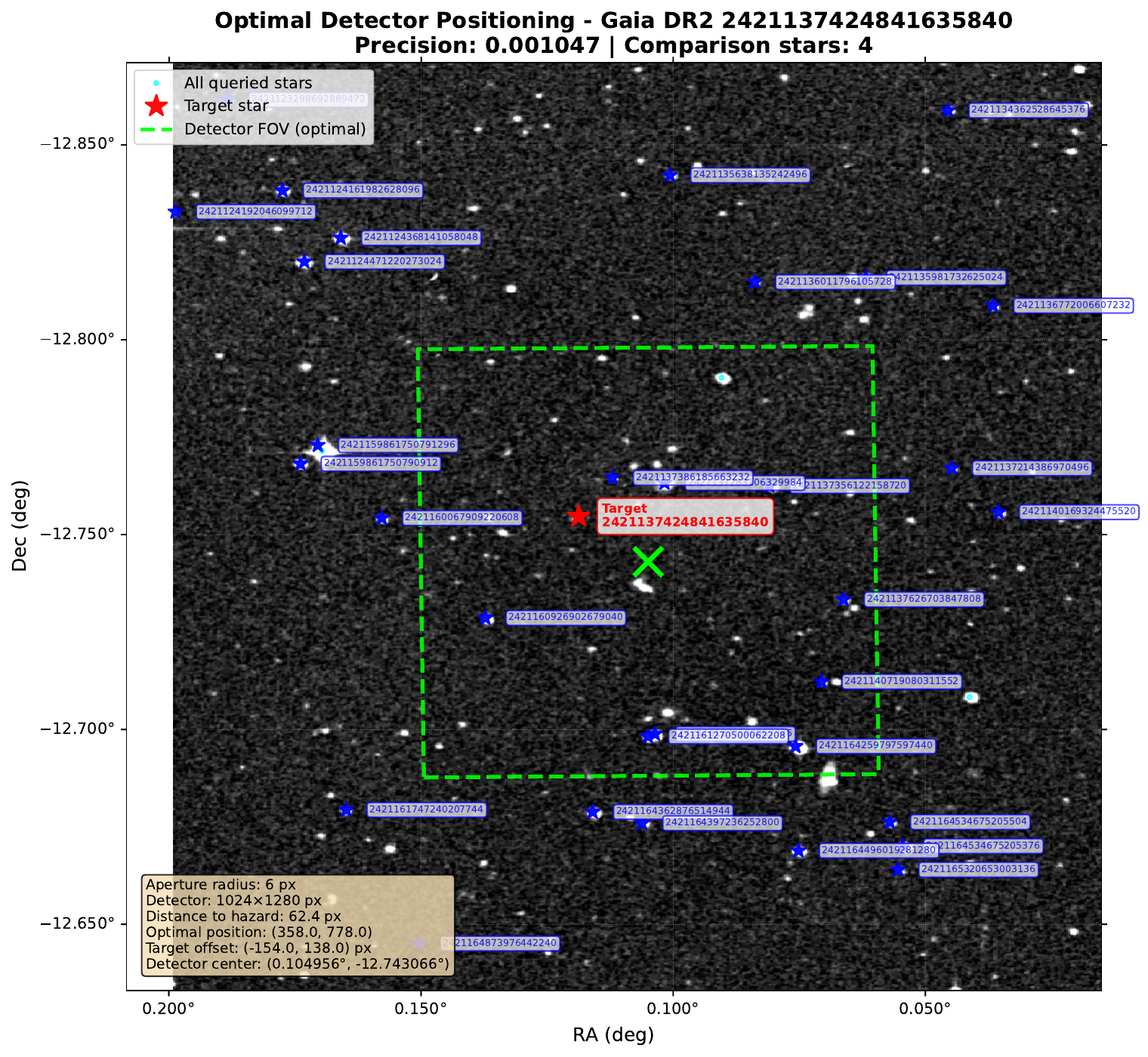}
    \caption{Output of the \texttt{goibniu} tool showing the optimal position on the detector to observe a target to maximise the photometric precision.}
    \end{subfigure}
\hfill
\begin{subfigure}[t]{.48\textwidth}
    \centering
    \includegraphics[height=7cm]{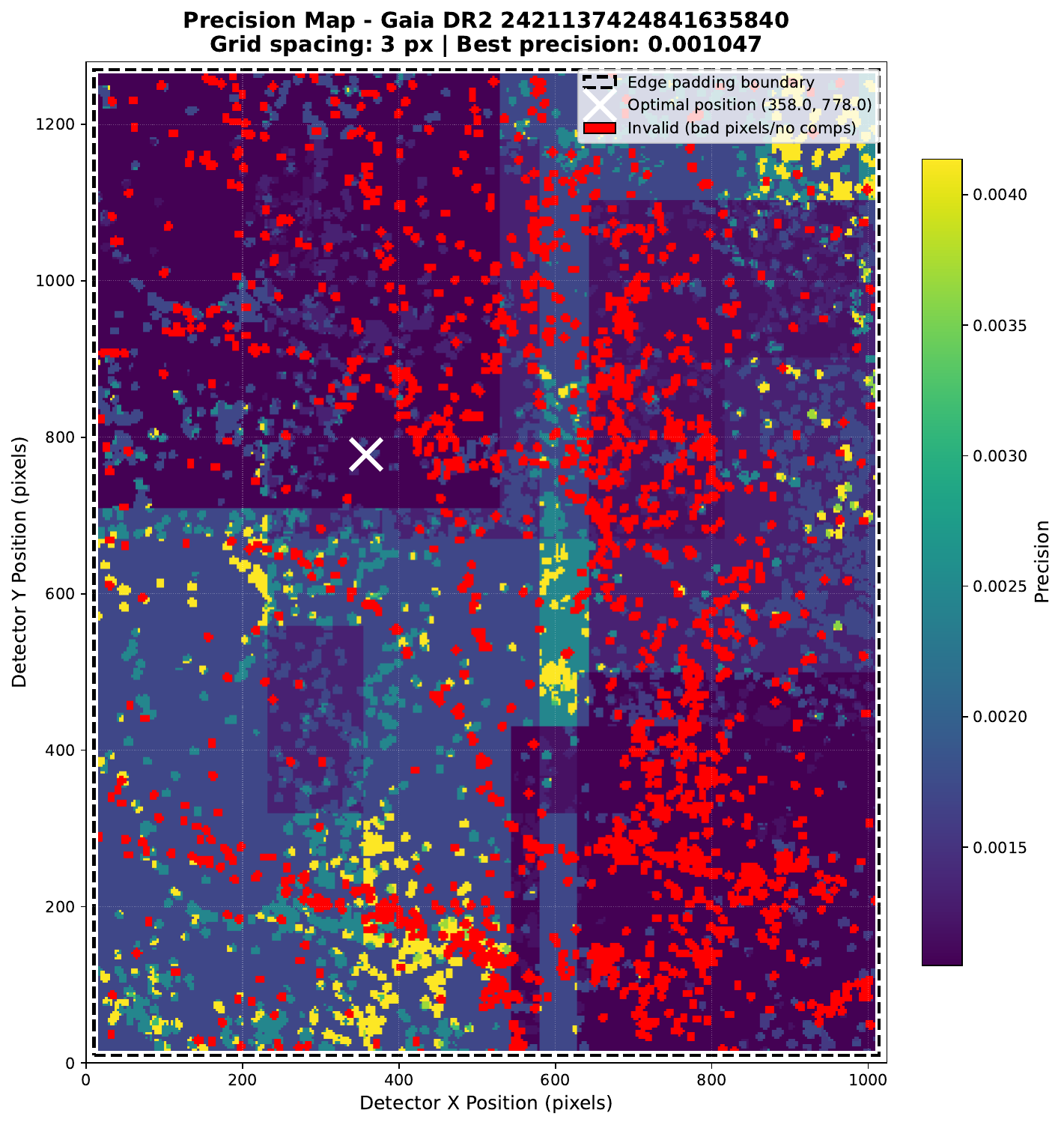}
    \caption{Map of SPIRIT's FOV generated by \texttt{goibniu} showing the expected precision obtained by placing the target at each location, given the bad pixel map and the available comparison stars. Positions excluded by bad pixels or a lack of comparison stars are shown in red.}
    \end{subfigure}
\caption{Example run of the \texttt{goibniu} tool for scheduling observations to optimise precision.}
\label{goibniu}
\end{figure*}

As discussed in Sections \ref{sec:methods} and \ref{sec:performance}, factors that still limit the precision of SPIRIT differential photometry are the number of comparison stars present in the field of view (FOV) of the target, and the effect of bad pixels on the detector. The \texttt{goibniu} tool (Hooton \& Janó-Muñoz, in prep.) computes the expected differential precision on a grid of positions covering the whole detector, based on the prospective comparison stars that lie within the FOV and do not coincide with bad pixels. For each grid position, the precision of each star in the FOV is interpolated from the precisions of stars of similar magnitude measured in the global analysis described in Section \ref{subsec:globalanalysis}. 
Following this, it returns the target XY position that yields the best differential precision while remaining as far as possible from bad pixels. This tool is now used when scheduling new SPIRIT targets, both to optimise the detector positioning and to discard targets for which good differential precision cannot be achieved. An example of an XY precision map produced by \texttt{goibniu} is shown in Figure \ref{goibniu}.

Future work will also characterise the effect of optimised, flexible aperture sizes on the photometric precision of the instrument, compared with the fixed apertures used in this work.

\subsection{Future expansion}
Another limiting factor when looking for exoplanets around UCDs is stellar activity. Even after optimising SPIRIT to potentially reach sub-Earth-radius detections, the amplitude of stellar variability can mask the transit signal. Our second case study, in Section \ref{subsec:BD}, shows the potential of monitoring UCDs in the NIR, particularly when combined with observations in visible bands. Simultaneous multi-band observations can disentangle the chromatic stellar signal and reveal largely wavelength-independent features in the data, such as transits. 

For this purpose, we developed DUET (DUal-band Explorer for Transits; \citealt[][]{pedersen2026duet}), a dual-channel imager for the SSO. A dichroic beam-splitter at 955\,nm directs visible light to the same model of deeply depleted silicon CCD as the Andor cameras used in this work, and near-infrared light to the same model of InGaAs CMOS detector as SPIRIT, enabling fully simultaneous photometry with bandpasses from 0.4 to 1.7\,$\text{\textmu}$m. 
The visible arm houses $g'$, $r'$, $i'$, $z'$, and $I+z$ filters, whereas the near-infrared arm contains $Y$, $YJ$, $J$, and $H$ filters. The NIR filters are optimised to suppress sensitivity to atmospheric PWV variability \citep[][]{Gourves2025}.
\section{Conclusions}
\label{sec:conclusion}

In this work, we present the on-sky performance of the InGaAs CMOS NIR detector SPIRIT over its first year of operation. SPIRIT is installed on the Callisto telescope at the SPECULOOS Southern Observatory. We use 328 visits of 73 targets during 2022 and 2023 to optimise calibration and characterise the limits of the detector's photometric precision. 

We detail our data reduction steps, which differ when moving from silicon-based CCDs to an InGaAs CMOS, and from the visible to the NIR. In particular, we describe our treatment of glow phenomena, which we correct through matched-exposure dark frames; our approach to the characteristic row-by-row readout pattern; and our handling of the $\sim$4,000 bad pixels (0.3\% of the array) through detection in dark and flat frames. We find that the effect of a single bad pixel falling within the target PSF is modest, worsening the 10-minute binned RMS by up to 1.2\% near the PSF centre, and conclude that while masking and interpolation allow us to meet our photometry goals, avoiding bad pixels altogether remains the best approach. We have applied our optimised calibration process to data taken during the current SPIRIT operation, from 2025 onwards.

We find that the measured precision agrees well with our theoretical noise model, reaching sub-ppt precision under good observing conditions. Based on this precision, we estimate that SPIRIT can detect the transits of Earth-sized planets around most of our targets, for a transit depth of four times the 10-minute binned RMS, noting that we are not accounting for the stellar variability of UCDs. We further quantify the effect of the change in operating temperature from \SI{-60}{\degreeCelsius} to \SI{-50}{\degreeCelsius}, adopted since 2025 to preserve the detector's longevity, and estimate an upper limit on the median increase in the 10-minute RMS of 1\%, comparable to the effect of a single bad pixel in the target PSF, rising to 68\% for the faintest targets. Measurements of the dark current in 2025 and 2026 indicate that 24--41\% of the dark current is thermally generated, allowing us to apply the conclusions of our global analysis to data taken after 2025, with the caveat that the faintest targets are the most affected.

SPIRIT's scientific capabilities are exemplified by two case studies. The simultaneous SPIRIT and CCD observation of a TRAPPIST-1\,h transit shows that the 10-minute binned precision is better with SPIRIT than with the CCD ($0.53$ versus $0.95$\,ppt), owing to its higher cadence and lower red noise: the $zYJ$ data are consistent with white noise, whereas the $I+z$ data show a red-noise contribution of $0.72$\,ppt, demonstrating the advantage of the $zYJ$ filter in minimising the effect of PWV. The monitoring of the fast-rotating dwarf LP~944-20 recovers its rotation period in the NIR, $P_{\rm rot} = 3.8411 \pm 0.0044$\,h, consistent with literature values obtained with TESS. We demonstrate SPIRIT's potential for activity monitoring and, when combined with visible-band observations, for disentangling wavelength-dependent stellar variability from largely wavelength-independent features such as transits.

Our results confirm that InGaAs infrared CMOS detectors are a cost-effective and competitive technology for ground-based transit photometry of cool stars. The non-cryogenic cooling requirements of InGaAs substantially lower installation and maintenance costs relative to HgCdTe or InSb detectors, without a prohibitive sacrifice in performance across the 0.81--1.33\,$\text{\textmu}$m range where UCDs emit most strongly. Building on this, we introduce tools such as \texttt{goibniu} to optimise target scheduling and detector positioning against the comparison-star and bad-pixel limitations identified in this work, as well as the forthcoming DUET dual-channel imager, which will extend our approach to fully simultaneous visible and NIR photometry from 0.4 to 1.7\,$\text{\textmu}$m. Together, these developments establish infrared CMOS photometry as a promising avenue for the detection and characterisation of temperate terrestrial planets around the coolest stars from the ground.

\section*{Acknowledgements}
Based on data collected by the SPECULOOS-South Observatory at the ESO Paranal Observatory in Chile. The ULiege's contribution to SPECULOOS has received funding from the European Research Council under the European Union's Seventh Framework Programme (FP/2007-2013) (grant Agreement n$^\circ$ 336480/SPECULOOS), from the Balzan Prize and Francqui Foundations, from the Belgian Scientific Research Foundation (F.R.S.-FNRS; grant n$^\circ$ T.0109.20), from the University of Liege, and from the ARC grant for Concerted Research Actions financed by the Wallonia-Brussels Federation. This work is supported by a grant from the Simons Foundation (PI Queloz, grant number 327127).
This research is in part funded by the European Union's Horizon 2020 research and innovation programme (grants agreements n$^{\circ}$ 803193/BEBOP), and from the Science and Technology Facilities Council (STFC; grant n$^\circ$ ST/S00193X/1, ST/W002582/1, and ST/Y001710/1). Funding for KB was provided by the European Union (ERC AdG SUBSTELLAR, GA 101054354).
This material is based upon work supported by the National Aeronautics and Space Administration under Agreement No.\ 80NSSC21K0593 for the program ``Alien Earths.''
The results reported herein benefited from collaborations and/or information exchange within NASA's Nexus for Exoplanet System Science (NExSS) research coordination network sponsored by NASA's Science Mission Directorate.
We disclose the use of AI tools in generating code for plotting, in cleaning code and in debugging. No AI was used to generate text in this work but was used for proof-reading and copy-editing. This work was supported by the University of Cambridge Harding Distinguished Postgraduate Scholars Programme.

\section*{Data Availability}

SPECULOOS data are publicly available at the European Southern Observatory (ESO) archive a year after acquisition.



\bibliographystyle{mnras}
\bibliography{refs}

@article{JanoMunoz2025,
   author = {C. Janó-Muñoz and M.~J. Hooton and P.~P. Pedersen and K Barkaoui and B.~V. Rackham and A.~J. Burgasser and F.~J. Pozuelos and K.~G. Stassun and D Queloz and A.~H.~M.~J. Triaud and C Ziegler and J.~M. Almenara and M Timmermans and X Bonfils and K.~A. Collins and B.~O. Demory and G Dransfield and M Ghachoui and M Gillon and E Jehin and A.~W. Mann and D Sebastian and S Thompson and J.~D. Twicken and J de Wit and S Zúñiga-Fernández},
   doi = {10.1093/mnras/staf946},
   journal = {\mnras},
   issue = {1},
   month = {7},
   pages = {630-641},
   title = {TOI-2407 b: a warm Neptune in the desert},
   volume = {541},
   year = {2025}
}

@ARTICLE{McCaughrean2002,
       author = {{McCaughrean}, M.~J. and {Scholz}, R.-D. and {Lodieu}, N.},
        title = "{Search for nearby stars among proper motion stars selected by optical-to-infrared photometry. II. Two late M dwarfs within 10 pc}",
      journal = {\aap},
         year = 2002,
        month = jul,
       volume = {390},
        pages = {L27-L30},
          doi = {10.1051/0004-6361:20020928},
archivePrefix = {arXiv},
       eprint = {astro-ph/0206378},
 primaryClass = {astro-ph},
       adsurl = {https://ui.adsabs.harvard.edu/abs/2002A&A...390L..27M}
}

@INPROCEEDINGS{Gourves2025,
       author = {{Gourv{\`e}s}, C. and {Strugarek}, A. and {Moutou}, C. and {Morin}, J. and {Petit}, P. and {Dyrek}, A. and {Breton}, S.~N. and {Garc{\'\i}a}, R.~A. and {Lanza}, A.~F. and {Mathur}, S. and {Santos}, A.~R.~G.},
        title = "{Towards the confirmation of a new ultra-short period planet orbiting a M-type star}",
    booktitle = {SF2A-2025: Semaine de l'Astrophysique Francaise},
         year = 2025,
       editor = {{Siebert}, A. and {Bailli{\'e}}, K. and {B{\'e}thermin}, M. and {Cantalloube}, F. and {Josselin}, E. and {Lagarde}, N. and {Malzac}, J. and {Richard}, J. and {Selliez}, L. and {Venot}, O.},
        month = dec,
        pages = {461-464},
       adsurl = {https://ui.adsabs.harvard.edu/abs/2025sf2a.conf..461G}
}

@ARTICLE{Pont2006,
       author = {{Pont}, Fr{\'e}d{\'e}ric and {Zucker}, Shay and {Queloz}, Didier},
        title = "{The effect of red noise on planetary transit detection}",
      journal = {\mnras},
         year = 2006,
        month = nov,
       volume = {373},
       number = {1},
        pages = {231-242},
          doi = {10.1111/j.1365-2966.2006.11012.x},
archivePrefix = {arXiv},
       eprint = {astro-ph/0608597},
 primaryClass = {astro-ph},
       adsurl = {https://ui.adsabs.harvard.edu/abs/2006MNRAS.373..231P}
}

@ARTICLE{Winn2008,
       author = {{Winn}, Joshua N. and {Holman}, Matthew J. and {Torres}, Guillermo and {McCullough}, Peter and {Johns-Krull}, Christopher and {Latham}, David W. and {Shporer}, Avi and {Mazeh}, Tsevi and {Garcia-Melendo}, Enrique and {Foote}, Cindy and {Esquerdo}, Gil and {Everett}, Mark},
        title = "{The Transit Light Curve Project. IX. Evidence for a Smaller Radius of the Exoplanet XO-3b}",
      journal = {\apj},
         year = 2008,
        month = aug,
       volume = {683},
       number = {2},
        pages = {1076-1084},
          doi = {10.1086/589737},
archivePrefix = {arXiv},
       eprint = {0804.4475},
 primaryClass = {astro-ph},
       adsurl = {https://ui.adsabs.harvard.edu/abs/2008ApJ...683.1076W}
}

@ARTICLE{Davoudi2024,
       author = {{Davoudi}, Fatemeh and {Rackham}, Benjamin V. and {Gillon}, Micha{\"e}l and {de Wit}, Julien and {Burgasser}, Adam J. and {Delrez}, Laetitia and {Iyer}, Aishwarya and {Ducrot}, Elsa},
        title = "{Updated Spectral Characteristics for the Ultracool Dwarf TRAPPIST-1}",
      journal = {\apjl},
         year = 2024,
        month = jul,
       volume = {970},
       number = {1},
          eid = {L4},
        pages = {L4},
          doi = {10.3847/2041-8213/ad5c6c},
       adsurl = {https://ui.adsabs.harvard.edu/abs/2024ApJ...970L...4D}
}

@ARTICLE{Lambier2025,
       author = {{Lambier}, Samantha and {Metchev}, Stanimir and {Miles-P{\'a}ez}, Paulo and {Moranta}, Leslie and {Wolfe}, Dakota and {Hales}, Joelene and {Martinovic}, Jeffrey},
        title = "{Rotation Periods of Candidate Single Late-M Dwarfs in TESS}",
      journal = {\aj},
         year = 2025,
        month = sep,
       volume = {170},
       number = {3},
          eid = {168},
        pages = {168},
          doi = {10.3847/1538-3881/adf195},
archivePrefix = {arXiv},
       eprint = {2507.10847},
 primaryClass = {astro-ph.SR},
       adsurl = {https://ui.adsabs.harvard.edu/abs/2025AJ....170..168L}
}

@ARTICLE{Tinney1998,
       author = {{Tinney}, C.~G. and {Reid}, I.~N.},
        title = "{High-resolution spectra of very low-mass stars}",
      journal = {\mnras},
         year = 1998,
        month = dec,
       volume = {301},
       number = {4},
        pages = {1031-1048},
          doi = {10.1046/j.1365-8711.1998.02079.x},
archivePrefix = {arXiv},
       eprint = {astro-ph/9806004},
 primaryClass = {astro-ph},
       adsurl = {https://ui.adsabs.harvard.edu/abs/1998MNRAS.301.1031T}
}

@ARTICLE{Clemens2007,
       author = {{Clemens}, D.~P. and {Sarcia}, D. and {Grabau}, A. and {Tollestrup}, E.~V. and {Buie}, M.~W. and {Dunham}, E. and {Taylor}, B.},
        title = "{Mimir: A Near-Infrared Wide-Field Imager, Spectrometer and Polarimeter}",
      journal = {\pasp},
         year = 2007,
        month = dec,
       volume = {119},
       number = {862},
        pages = {1385-1402},
          doi = {10.1086/524775},
       adsurl = {https://ui.adsabs.harvard.edu/abs/2007PASP..119.1385C}
}

@ARTICLE{Tamburo2022,
       author = {{Tamburo}, Patrick and {Muirhead}, Philip S. and {McCarthy}, Allison M. and {Hart}, Murdock and {Gracia}, David and {Vos}, Johanna M. and {Bardalez Gagliuffi}, Daniella C. and {Faherty}, Jacqueline and {Theissen}, Christopher and {Agol}, Eric and {Skinner}, Julie N. and {Sagear}, Sheila},
        title = "{The Perkins INfrared Exosatellite Survey (PINES) I. Survey Overview, Reduction Pipeline, and Early Results}",
      journal = {\aj},
         year = 2022,
        month = jun,
       volume = {163},
       number = {6},
          eid = {253},
        pages = {253},
          doi = {10.3847/1538-3881/ac64aa},
archivePrefix = {arXiv},
       eprint = {2201.01794},
 primaryClass = {astro-ph.EP},
       adsurl = {https://ui.adsabs.harvard.edu/abs/2022AJ....163..253T}
}

@article{Palle2025,
   author = {Enric Palle and Katia Biazzo and Emeline Bolmont and Paul Mollière and Katja Poppenhaeger and Jayne Birkby and Matteo Brogi and Gael Chauvin and Andrea Chiavassa and Jens Hoeijmakers and Emmanuel Lellouch and Christophe Lovis and Roberto Maiolino and Lisa Nortmann and Hannu Parviainen and Lorenzo Pino and Martin Turbet and Jesse Weder and Simon Albrecht and Simone Antoniucci and Susana C Barros and Andre Beaudoin and Bjorn Benneke and Isabelle Boisse and Aldo S Bonomo and Francesco Borsa and Alexis Brandeker and Wolfgang Brandner and Lars A Buchhave and Anne-Laure Cheffot and Robin Deborde and Florian Debras and Rene Doyon and Paolo Di Marcantonio and Paolo Giacobbe and Jonay I González Hernández and Ravit Helled and Laura Kreidberg and Pedro Machado and Jesus Maldonado and Alessandro Marconi and B.~L. Canto Martins and Adriano Miceli and Christoph Mordasini and Mamadou N'Diaye and Andrzej Niedzielski and Brunella Nisini and Livia Origlia and Celine Peroux and Alexander G.~M. Pietrow and Enrico Pinna and Emily Rauscher and Sabine Reffert and Cristina Rodr\'\iguez-López and Philippe Rousselot and Nicoletta Sanna and Nuno C Santos and Adrien Simonnin and Alejandro Suárez Mascareño and Alessio Zanutta and Maria Rosa Zapatero-Osorio and Mathias Zechmeister},
   doi = {10.1007/s10686-025-10000-4},
   issue = {3},
   journal = {Experimental Astronomy},
   month = {6},
   pages = {29},
   title = {Ground-breaking exoplanet science with the ANDES spectrograph at the ELT},
   volume = {59},
   year = {2025}
}

@ARTICLE{Gardner2023,
       author = {{Gardner}, Jonathan P. and {Mather}, John C. and {Abbott}, Randy and {Abell}, James S. and {Abernathy}, Mark and {Abney}, Faith E. and {Abraham}, John G. and {Abraham}, Roberto and {Abul-Huda}, Yasin M. and {Acton}, Scott and {Adams}, Cynthia K. and {Adams}, Evan and {Adler}, David S. and {Adriaensen}, Maarten and {Aguilar}, Jonathan Albert and {Ahmed}, Mansoor and {Ahmed}, Nasif S. and {Ahmed}, Tanjira and {Albat}, R{\"u}deger and {Albert}, Lo{\"\i}c and {Alberts}, Stacey and {Aldridge}, David and {Allen}, Mary Marsha and {Allen}, Shaune S. and {Altenburg}, Martin and {Altunc}, Serhat and {Alvarez}, Jose Lorenzo and {{\'A}lvarez-M{\'a}rquez}, Javier and {Alves de Oliveira}, Catarina and {Ambrose}, Leslie L. and {Anandakrishnan}, Satya M. and {Andersen}, Gregory C. and {Anderson}, Harry James and {Anderson}, Jay and {Anderson}, Kristen and {Anderson}, Sara M. and {Aprea}, Julio and {Archer}, Benita J. and {Arenberg}, Jonathan W. and {Argyriou}, Ioannis and {Arribas}, Santiago and {Artigau}, {\'E}tienne and {Arvai}, Amanda Rose and {Atcheson}, Paul and {Atkinson}, Charles B. and {Averbukh}, Jesse and {Aymergen}, Cagatay and {Bacinski}, John J. and {Baggett}, Wayne E. and {Bagnasco}, Giorgio and {Baker}, Lynn L. and {Balzano}, Vicki Ann and {Banks}, Kimberly A. and {Baran}, David A. and {Barker}, Elizabeth A. and {Barrett}, Larry K. and {Barringer}, Bruce O. and {Barto}, Allison and {Bast}, William and {Baudoz}, Pierre and {Baum}, Stefi and {Beatty}, Thomas G. and {Beaulieu}, Mathilde and {Bechtold}, Kathryn and {Beck}, Tracy and {Beddard}, Megan M. and {Beichman}, Charles and {Bellagama}, Larry and {Bely}, Pierre and {Berger}, Timothy W. and {Bergeron}, Louis E. and {Bernier}, Antoine-Darveau and {Bertch}, Maria D. and {Beskow}, Charlotte and {Betz}, Laura E. and {Biagetti}, Carl P. and {Birkmann}, Stephan and {Bjorklund}, Kurt F. and {Blackwood}, James D. and {Blazek}, Ronald Paul and {Blossfeld}, Stephen and {Bluth}, Marcel and {Boccaletti}, Anthony and {Boegner}, Jr., Martin E. and {Bohlin}, Ralph C. and {Boia}, John Joseph and {B{\"o}ker}, Torsten and {Bonaventura}, N. and {Bond}, Nicholas A. and {Bosley}, Kari Ann and {Boucarut}, Rene A. and {Bouchet}, Patrice and {Bouwman}, Jeroen and {Bower}, Gary and {Bowers}, Ariel S. and {Bowers}, Charles W. and {Boyce}, Leslye A. and {Boyer}, Christine T. and {Boyer}, Martha L. and {Boyer}, Michael and {Boyer}, Robert and {Bradley}, Larry D. and {Brady}, Gregory R. and {Brandl}, Bernhard R. and {Brannen}, Judith L. and {Breda}, David and {Bremmer}, Harold G. and {Brennan}, David and {Bresnahan}, Pamela A. and {Bright}, Stacey N. and {Broiles}, Brian J. and {Bromenschenkel}, Asa and {Brooks}, Brian H. and {Brooks}, Keira J. and {Brown}, Bob and {Brown}, Bruce and {Brown}, Thomas M. and {Bruce}, Barry W. and {Bryson}, Jonathan G. and {Bujanda}, Edwin D. and {Bullock}, Blake M. and {Bunker}, A.~J. and {Bureo}, Rafael and {Burt}, Irving J. and {Bush}, James Aaron and {Bushouse}, Howard A. and {Bussman}, Marie C. and {Cabaud}, Olivier and {Cale}, Steven and {Calhoon}, Charles D. and {Calvani}, Humberto and {Canipe}, Alicia M. and {Caputo}, Francis M. and {Cara}, Mihai and {Carey}, Larkin and {Case}, Michael Eli and {Cesari}, Thaddeus and {Cetorelli}, Lee D. and {Chance}, Don R. and {Chandler}, Lynn and {Chaney}, Dave and {Chapman}, George N. and {Charlot}, S. and {Chayer}, Pierre and {Cheezum}, Jeffrey I. and {Chen}, Bin and {Chen}, Christine H. and {Cherinka}, Brian and {Chichester}, Sarah C. and {Chilton}, Zachary S. and {Chittiraibalan}, Dharini and {Clampin}, Mark and {Clark}, Charles R. and {Clark}, Kerry W. and {Clark}, Stephanie M. and {Claybrooks}, Edward E. and {Cleveland}, Keith A. and {Cohen}, Andrew L. and {Cohen}, Lester M. and {Col{\'o}n}, Knicole D. and {Coleman}, Benee L. and {Colina}, Luis and {Comber}, Brian J. and {Comeau}, Thomas M. and {Comer}, Thomas and {Conde Reis}, Alain and {Connolly}, Dennis C. and {Conroy}, Kyle E. and {Contos}, Adam R. and {Contreras}, James and {Cook}, Neil J. and {Cooper}, James L. and {Cooper}, Rachel Aviva and {Correia}, Michael F. and {Correnti}, Matteo and {Cossou}, Christophe and {Costanza}, Brian F. and {Coulais}, Alain and {Cox}, Colin R. and {Coyle}, Ray T. and {Cracraft}, Misty M. and {Crew}, Keith A. and {Curtis}, Gary J. and {Cusveller}, Bianca and {Da Costa Maciel}, Cleyciane and {Dailey}, Christopher T. and {Daugeron}, Fr{\'e}d{\'e}ric and {Davidson}, Greg S. and {Davies}, James E. and {Davis}, Katherine Anne and {Davis}, Michael S. and {Day}, Ratna and {de Chambure}, Daniel and {de Jong}, Pauline and {De Marchi}, Guido and {Dean}, Bruce H. and {Decker}, John E. and {Delisa}, Amy S. and {Dell}, Lawrence C. and {Dellagatta}, Gail},
        title = "{The James Webb Space Telescope Mission}",
      journal = {\pasp},
         year = 2023,
        month = jun,
       volume = {135},
       number = {1048},
          eid = {068001},
        pages = {068001},
          doi = {10.1088/1538-3873/acd1b5},
archivePrefix = {arXiv},
       eprint = {2304.04869},
 primaryClass = {astro-ph.IM},
       adsurl = {https://ui.adsabs.harvard.edu/abs/2023PASP..135f8001G}
}

@inproceedings{SaintEx,
   author = {Laurence Sabin and Yilen Gómez Maqueo Chew and Brice-Olivier Demory and Romina Petrucci and Saint-Ex Consortium},
   doi = {10.5281/zenodo.1479857},
   booktitle = {20th Cambridge Workshop on Cool Stars, Stellar Systems and the Sun},
   month = {7},
   pages = {59},
   title = {SAINT-EX: Searching for exoplanets from the San Pedro Mártir Observatory},
   year = {2018}
}

@ARTICLE{Triaud2023,
       author = {{Triaud}, Amaury H.~M.~J. and {Dransfield}, Georgina and {Kagetani}, Taiki and {Timmermans}, Mathilde and {Narita}, Norio and {Barkaoui}, Khalid and {Hirano}, Teruyuki and {Rackham}, Benjamin V. and {Mori}, Mayuko and {Baycroft}, Thomas and {Benkhaldoun}, Zouhair and {Burgasser}, Adam J. and {Caldwell}, Douglas A. and {Collins}, Karen A. and {Davis}, Yasmin T. and {Delrez}, Laetitia and {Demory}, Brice-Oliver and {Ducrot}, Elsa and {Fukui}, Akihiko and {Mu{\~n}oz}, Cl{\`a}udia Jano and {Jehin}, Emmanu{\"e}l and {Garc{\'\i}a}, Lionel J. and {Ghachoui}, Mourad and {Gillon}, Micha{\"e}l and {Chew}, Yilen G{\'o}mez Maqueo and {Hooton}, Matthew J. and {Ikoma}, Masahiro and {Kawauchi}, Kiyoe and {Kotani}, Takayuki and {Levine}, Alan M. and {Pall{\'e}}, Enric and {Pedersen}, Peter P. and {Pozuelos}, Francisco J. and {Queloz}, Didier and {Scutt}, Owen J. and {Seager}, Sara and {Sebastian}, Daniel and {Tamura}, Motohide and {Thompson}, Samantha and {Watanabe}, Noriharu and {de Wit}, Julien and {Winn}, Joshua N. and {Z{\'u}{\~n}iga-Fern{\'a}ndez}, Sebasti{\'a}n},
        title = "{An M dwarf accompanied by a close-in giant orbiter with SPECULOOS}",
      journal = {\mnras},
         year = 2023,
        month = oct,
       volume = {525},
       number = {1},
        pages = {L98-L104},
          doi = {10.1093/mnrasl/slad097},
archivePrefix = {arXiv},
       eprint = {2308.02253},
 primaryClass = {astro-ph.EP},
       adsurl = {https://ui.adsabs.harvard.edu/abs/2023MNRAS.525L..98T}
}

@article{Dransfield2025,
   author = {G Dransfield and M Timmermans and D Sebastian and B.~V. Rackham and A Burgasser and K Barkaoui and A.~H.~M.~J. Triaud and M Gillon and J.~M. Almenara and S.~L. Casewell and K.~A. Collins and A Fukui and C Jano-Munoz and S Kanodia and N Narita and E Palle and M.~G. Scott and A Soubkiou and A Stokholm and J Audenaert and G Á Bakos and Y Beletsky and Z.~L. de Beurs and Z Benkhaldoun and A Burdanov and R.~P. Butler and D Caldwell and J.~D. Crane and Y.~T. Davis and B.~O. Demory and E Ducrot and Y Gómez Maqueo Chew and M Gachaoui and J.~D. Hartman and M.~J. Hooton and E Jehin and S Mercier and F Murgas and C Murray and P.~P. Pedersen and F.~J. Pozuelos and M Rice and G Ross and S.~A. Shectman and E Softich and M Tala Pinto and A.~M. Vanderburg and J Villasenor and J de Wit and S Zúñiga-Fernández},
   doi = {10.48550/arXiv.2510.11528},
   journal = {arXiv e-prints},
   month = {10},
   pages = {arXiv:2510.11528},
   title = {MANGOS II: Five new giant planets orbiting low-mass stars},
   year = {2025}
}

@article{Gillon2017,
   author = {Michaël Gillon and Amaury H.~M.~J. Triaud and Brice-Olivier Demory and Emmanuël Jehin and Eric Agol and Katherine M Deck and Susan M Lederer and Julien de Wit and Artem Burdanov and James G Ingalls and Emeline Bolmont and Jeremy Leconte and Sean N Raymond and Franck Selsis and Martin Turbet and Khalid Barkaoui and Adam Burgasser and Matthew R Burleigh and Sean J Carey and Aleksander Chaushev and Chris M Copperwheat and Laetitia Delrez and Catarina S Fernandes and Daniel L Holdsworth and Enrico J Kotze and Valérie Van Grootel and Yaseen Almleaky and Zouhair Benkhaldoun and Pierre Magain and Didier Queloz},
   doi = {10.1038/nature21360},
   journal = {Nature},
   issue = {7642},
   month = {2},
   pages = {456-460},
   title = {Seven temperate terrestrial planets around the nearby ultracool dwarf star TRAPPIST-1},
   volume = {542},
   year = {2017}
}

@INPROCEEDINGS{Delrez2018,
       author = {{Delrez}, Laetitia and {Gillon}, Micha{\"e}l. and {Queloz}, Didier and {Demory}, Brice-Olivier and {Almleaky}, Yaseen and {de Wit}, Julien and {Jehin}, Emmanu{\"e}l. and {Triaud}, Amaury H.~M.~J. and {Barkaoui}, Khalid and {Burdanov}, Artem and {Burgasser}, Adam J. and {Ducrot}, Elsa and {McCormac}, James and {Murray}, Catriona and {Silva Fernandes}, Catarina and {Sohy}, Sandrine and {Thompson}, Samantha J. and {Van Grootel}, Val{\'e}rie and {Alonso}, Roi and {Benkhaldoun}, Zouhair and {Rebolo}, Rafael},
        title = "{SPECULOOS: a network of robotic telescopes to hunt for terrestrial planets around the nearest ultracool dwarfs}",
    booktitle = {Ground-based and Airborne Telescopes VII},
         year = 2018,
       editor = {{Marshall}, Heather K. and {Spyromilio}, Jason},
       series = {Society of Photo-Optical Instrumentation Engineers (SPIE) Conference Series},
       volume = {10700},
        month = jul,
          eid = {107001I},
        pages = {107001I},
          doi = {10.1117/12.2312475},
archivePrefix = {arXiv},
       eprint = {1806.11205},
 primaryClass = {astro-ph.IM},
       adsurl = {https://ui.adsabs.harvard.edu/abs/2018SPIE10700E..1ID}
}

@ARTICLE{Sebastian2021,
       author = {{Sebastian}, D. and {Gillon}, M. and {Ducrot}, E. and {Pozuelos}, F.~J. and {Garcia}, L.~J. and {G{\"u}nther}, M.~N. and {Delrez}, L. and {Queloz}, D. and {Demory}, B.~O. and {Triaud}, A.~H.~M.~J. and {Burgasser}, A. and {de Wit}, J. and {Burdanov}, A. and {Dransfield}, G. and {Jehin}, E. and {McCormac}, J. and {Murray}, C.~A. and {Niraula}, P. and {Pedersen}, P.~P. and {Rackham}, B.~V. and {Sohy}, S. and {Thompson}, S. and {Van Grootel}, V.},
        title = "{SPECULOOS: Ultracool dwarf transit survey. Target list and strategy}",
      journal = {\aap},
         year = 2021,
        month = jan,
       volume = {645},
          eid = {A100},
        pages = {A100},
          doi = {10.1051/0004-6361/202038827},
archivePrefix = {arXiv},
       eprint = {2011.02069},
 primaryClass = {astro-ph.EP},
       adsurl = {https://ui.adsabs.harvard.edu/abs/2021A&A...645A.100S}
}

@INCOLLECTION{Burdanov2018,
       author = {{Burdanov}, Artem and {Delrez}, Laetitia and {Gillon}, Micha{\"e}l and {Jehin}, Emmanu{\"e}l},
        title = "{SPECULOOS Exoplanet Search and Its Prototype on TRAPPIST}",
    booktitle = {Handbook of Exoplanets},
         year = 2018,
       editor = {{Deeg}, Hans J. and {Belmonte}, Juan Antonio},
          eid = {130},
        pages = {130},
          doi = {10.1007/978-3-319-55333-7_130},
       adsurl = {https://ui.adsabs.harvard.edu/abs/2018haex.bookE.130B}
}

@ARTICLE{Murray2020,
       author = {{Murray}, C.~A. and {Delrez}, L. and {Pedersen}, P.~P. and {Queloz}, D. and {Gillon}, M. and {Burdanov}, A. and {Ducrot}, E. and {Garcia}, L.~J. and {Lienhard}, F. and {Demory}, B.~O. and {Jehin}, E. and {McCormac}, J. and {Sebastian}, D. and {Sohy}, S. and {Thompson}, S.~J. and {Triaud}, A.~H.~M.~J. and {Van Grootel}, V. and {G{\"u}nther}, M.~N. and {Huang}, C.~X.},
        title = "{Photometry and performance of SPECULOOS-South}",
      journal = {\mnras},
         year = 2020,
        month = jun,
       volume = {495},
       number = {2},
        pages = {2446-2457},
          doi = {10.1093/mnras/staa1283},
archivePrefix = {arXiv},
       eprint = {2005.02423},
 primaryClass = {astro-ph.EP},
       adsurl = {https://ui.adsabs.harvard.edu/abs/2020MNRAS.495.2446M}
}

@INPROCEEDINGS{Gillon2011,
       author = {{Gillon}, M. and {Jehin}, E. and {Magain}, P. and {Chantry}, V. and {Hutsem{\'e}kers}, D. and {Manfroid}, J. and {Queloz}, D. and {Udry}, S.},
        title = "{TRAPPIST: a robotic telescope dedicated to the study of planetary systems}",
    booktitle = {European Physical Journal Web of Conferences},
         year = 2011,
       series = {European Physical Journal Web of Conferences},
       volume = {11},
        month = feb,
          eid = {06002},
        pages = {06002},
          doi = {10.1051/epjconf/20101106002},
archivePrefix = {arXiv},
       eprint = {1101.5807},
 primaryClass = {astro-ph.EP},
       adsurl = {https://ui.adsabs.harvard.edu/abs/2011EPJWC..1106002G}
}

@ARTICLE{Lomb,
       author = {{Lomb}, N.~R.},
        title = "{Least-Squares Frequency Analysis of Unequally Spaced Data}",
      journal = {\apss},
         year = 1976,
        month = feb,
       volume = {39},
       number = {2},
        pages = {447-462},
          doi = {10.1007/BF00648343},
       adsurl = {https://ui.adsabs.harvard.edu/abs/1976Ap&SS..39..447L}
}

@ARTICLE{Stellingwerf1978,
       author = {{Stellingwerf}, R.~F.},
        title = "{Period determination using phase dispersion minimization.}",
      journal = {\apj},
         year = 1978,
        month = sep,
       volume = {224},
        pages = {953-960},
          doi = {10.1086/156444},
       adsurl = {https://ui.adsabs.harvard.edu/abs/1978ApJ...224..953S}
}

@ARTICLE{ESOsky,
       author = {{Noll}, S. and {Kausch}, W. and {Barden}, M. and {Jones}, A.~M. and {Szyszka}, C. and {Kimeswenger}, S. and {Vinther}, J.},
        title = "{An atmospheric radiation model for Cerro Paranal. I. The optical spectral range}",
      journal = {\aap},
         year = 2012,
        month = jul,
       volume = {543},
          eid = {A92},
        pages = {A92},
          doi = {10.1051/0004-6361/201219040},
archivePrefix = {arXiv},
       eprint = {1205.2003},
 primaryClass = {astro-ph.IM},
       adsurl = {https://ui.adsabs.harvard.edu/abs/2012A&A...543A..92N}
}

@ARTICLE{Harding2013,
       author = {{Harding}, L.~K. and {Hallinan}, G. and {Boyle}, R.~P. and {Golden}, A. and {Singh}, Navtej and {Sheehan}, B. and {Zavala}, R.~T. and {Butler}, R.~F.},
        title = "{Periodic Optical Variability of Radio-detected Ultracool Dwarfs}",
      journal = {\apj},
         year = 2013,
        month = dec,
       volume = {779},
       number = {2},
          eid = {101},
        pages = {101},
          doi = {10.1088/0004-637X/779/2/101},
archivePrefix = {arXiv},
       eprint = {1310.1367},
 primaryClass = {astro-ph.SR},
       adsurl = {https://ui.adsabs.harvard.edu/abs/2013ApJ...779..101H}
}

@ARTICLE{Popowicz2013,
       author = {{Popowicz}, A. and {Kurek}, A.~R. and {Filus}, Z.},
        title = "{Bad Pixel Modified Interpolation for Astronomical Images}",
      journal = {\pasp},
         year = 2013,
        month = sep,
       volume = {125},
       number = {931},
        pages = {1119},
          doi = {10.1086/673179},
archivePrefix = {arXiv},
       eprint = {1309.4224},
 primaryClass = {astro-ph.IM},
       adsurl = {https://ui.adsabs.harvard.edu/abs/2013PASP..125.1119P}
}

@ARTICLE{Scargle,
       author = {{Scargle}, J.~D.},
        title = "{Studies in astronomical time series analysis. II. Statistical aspects of spectral analysis of unevenly spaced data.}",
      journal = {\apj},
         year = 1982,
        month = dec,
       volume = {263},
        pages = {835-853},
          doi = {10.1086/160554},
       adsurl = {https://ui.adsabs.harvard.edu/abs/1982ApJ...263..835S}
}

@ARTICLE{Tam1984,
  author={Tam, S. and Chenming Hu},
  journal={IEEE Transactions on Electron Devices}, 
  title={Hot-electron-induced photon and photocarrier generation in Silicon MOSFET's}, 
  year={1984},
  volume={31},
  number={9},
  pages={1264-1273},
  doi={10.1109/T-ED.1984.21698}}

@ARTICLE{Pedersen2023,
       author = {{Pedersen}, Peter P. and {Murray}, C.~A. and {Queloz}, D. and {Gillon}, M. and {Demory}, B.~O. and {Triaud}, A.~H.~M.~J. and {de Wit}, J. and {Delrez}, L. and {Dransfield}, G. and {Ducrot}, E. and {Garcia}, L.~J. and {G{\'o}mez Maqueo Chew}, Y. and {G{\"u}nther}, M.~N. and {Jehin}, E. and {McCormac}, J. and {Niraula}, P. and {Pozuelos}, F.~J. and {Rackham}, B.~V. and {Schanche}, N. and {Sebastian}, D. and {Thompson}, S.~J. and {Timmermans}, M. and {Wells}, R.},
        title = "{Precise near-infrared photometry, accounting for precipitable water vapour at SPECULOOS Southern Observatory}",
      journal = {\mnras},
         year = 2023,
        month = jan,
       volume = {518},
       number = {2},
        pages = {2661-2670},
          doi = {10.1093/mnras/stac3154},
archivePrefix = {arXiv},
       eprint = {2211.00156},
 primaryClass = {astro-ph.IM},
       adsurl = {https://ui.adsabs.harvard.edu/abs/2023MNRAS.518.2661P}
}

@ARTICLE{Layden2025,
       author = {{Layden}, Christopher and {Juneau}, Jill and {Pettersson}, Gustav and {Lourie}, Nathan and {Schneider}, Benjamin and {LaMarr}, Beverly and {Elio Angile}, Francesco and {Farag}, Fadi and {Luo}, Michelle and {Ong}, Zhi Zheng and {Furesz}, Gabor},
        title = "{Characterization of the Teledyne COSMOS camera: a large format CMOS image sensor for astronomy}",
      journal = {Journal of Astronomical Telescopes, Instruments, and Systems},
         year = 2025,
        month = apr,
       volume = {11},
          eid = {026003},
        pages = {026003},
          doi = {10.1117/1.JATIS.11.2.026003},
archivePrefix = {arXiv},
       eprint = {2502.00101},
 primaryClass = {astro-ph.IM},
       adsurl = {https://ui.adsabs.harvard.edu/abs/2025JATIS..11b6003L}
}

@ARTICLE{Regan2020,
       author = {{Regan}, Michael W. and {Bergeron}, Louis E.},
        title = "{Zero dark current in H2RG detectors: it is all multiplexer glow}",
      journal = {Journal of Astronomical Telescopes, Instruments, and Systems},
         year = 2020,
        month = jan,
       volume = {6},
          eid = {016001},
        pages = {016001},
          doi = {10.1117/1.JATIS.6.1.016001},
       adsurl = {https://ui.adsabs.harvard.edu/abs/2020JATIS...6a6001R}
}

@ARTICLE{Fetherolf2023,
       author = {{Fetherolf}, Tara and {Pepper}, Joshua and {Simpson}, Emilie and {Kane}, Stephen R. and {Mo{\v{c}}nik}, Teo and {English}, John Edward and {Antoci}, Victoria and {Huber}, Daniel and {Jenkins}, Jon M. and {Stassun}, Keivan and {Twicken}, Joseph D. and {Vanderspek}, Roland and {Winn}, Joshua N.},
        title = "{Variability Catalog of Stars Observed during the TESS Prime Mission}",
      journal = {\apjs},
         year = 2023,
        month = sep,
       volume = {268},
       number = {1},
          eid = {4},
        pages = {4},
          doi = {10.3847/1538-4365/acdee5},
archivePrefix = {arXiv},
       eprint = {2208.11721},
 primaryClass = {astro-ph.SR},
       adsurl = {https://ui.adsabs.harvard.edu/abs/2023ApJS..268....4F}
}

@INPROCEEDINGS{SPIRIT,
       author = {{Pedersen}, Peter P. and {Queloz}, Didier and {Garcia}, Lionel and {Schacke}, Yannick and {Delrez}, Laetitia and {Demory}, Brice-Olivier and {Ducrot}, Elsa and {Dransfield}, Georgina and {Gillon}, Michael and {Hooton}, Matthew J. and {Jan{\'o}-Mu{\~n}oz}, Cl{\`a}udia and {Jehin}, Emmanu{\"e}l. and {Sebastian}, Daniel and {Timmermans}, Mathilde and {Thompson}, Samantha and {Triaud}, Amaury H.~M.~J. and {de Wit}, Julien and {Z{\'u}{\~n}iga-Fern{\'a}ndez}, Sebasti{\'a}n.},
        title = "{Infrared photometry with InGaAs detectors: first light with SPECULOOS}",
    booktitle = {Ground-based and Airborne Instrumentation for Astronomy X},
         year = 2024,
       editor = {{Bryant}, Julia J. and {Motohara}, Kentaro and {Vernet}, Jo{\"e}l. R.~D.},
       series = {Society of Photo-Optical Instrumentation Engineers (SPIE) Conference Series},
       volume = {13096},
        month = jul,
          eid = {130963X},
        pages = {130963X},
          doi = {10.1117/12.3018320},
       adsurl = {https://ui.adsabs.harvard.edu/abs/2024SPIE13096E..3XP}
}

@ARTICLE{CCD,
       author = {{Boyle}, Willard and {Smith}, George},
        title = "{Charge Coupled Semiconductor Devices}",
      journal = {Bell System Technical Journal},
         year = 1970,
        month = apr,
       volume = {49},
        pages = {587-593},
          doi = {10.1002/j.1538-7305.1970.tb01790.x},
       adsurl = {https://ui.adsabs.harvard.edu/abs/1970BSTJ...49..587B}
}

@ARTICLE{CMOS,
       author = {{Fossum}, E.~R.},
        title = "{CMOS image sensors: electronic camera-on-a-chip}",
      journal = {IEEE Transactions on Electron Devices},
         year = 1997,
        month = oct,
       volume = {44},
       number = {10},
        pages = {1689-1698},
          doi = {10.1109/16.628824},
       adsurl = {https://ui.adsabs.harvard.edu/abs/1997ITED...44.1689F}
}

@ARTICLE{Alarcon_2023,
       author = {{Alarcon}, Miguel R. and {Licandro}, Javier and {Serra-Ricart}, Miquel and {Joven}, Enrique and {Gaitan}, Vicens and {de Sousa}, Rebeca},
        title = "{Scientific CMOS Sensors in Astronomy: IMX455 and IMX411}",
      journal = {\pasp},
         year = 2023,
        month = may,
       volume = {135},
       number = {1047},
          eid = {055001},
        pages = {055001},
          doi = {10.1088/1538-3873/acd04a},
archivePrefix = {arXiv},
       eprint = {2302.03700},
 primaryClass = {astro-ph.IM},
       adsurl = {https://ui.adsabs.harvard.edu/abs/2023PASP..135e5001A}
}

@ARTICLE{BIC,
       author = {{Schwarz}, Gideon},
        title = "{Estimating the Dimension of a Model}",
      journal = {Annals of Statistics},
         year = 1978,
        month = jul,
       volume = {6},
       number = {2},
        pages = {461-464},
       adsurl = {https://ui.adsabs.harvard.edu/abs/1978AnSta...6..461S}
}

@INPROCEEDINGS{Bonfils2015,
       author = {{Bonfils}, X. and {Almenara}, J.~M. and {Jocou}, L. and {Wunsche}, A. and {Kern}, P. and {Delboulb{\'e}}, A. and {Delfosse}, X. and {Feautrier}, P. and {Forveille}, T. and {Gluck}, L. and {Lafrasse}, S. and {Magnard}, Y. and {Maurel}, D. and {Moulin}, T. and {Murgas}, F. and {Rabou}, P. and {Rochat}, S. and {Roux}, A. and {Stadler}, E.},
        title = "{ExTrA: Exoplanets in transit and their atmospheres}",
    booktitle = {Techniques and Instrumentation for Detection of Exoplanets VII},
         year = 2015,
       editor = {{Shaklan}, Stuart},
       series = {Society of Photo-Optical Instrumentation Engineers (SPIE) Conference Series},
       volume = {9605},
        month = sep,
          eid = {96051L},
        pages = {96051L},
          doi = {10.1117/12.2186999},
archivePrefix = {arXiv},
       eprint = {1508.06601},
 primaryClass = {astro-ph.IM},
       adsurl = {https://ui.adsabs.harvard.edu/abs/2015SPIE.9605E..1LB}
}

@ARTICLE{batman,
       author = {{Kreidberg}, Laura},
        title = "{batman: BAsic Transit Model cAlculatioN in Python}",
      journal = {\pasp},
         year = 2015,
        month = nov,
       volume = {127},
       number = {957},
        pages = {1161},
          doi = {10.1086/683602},
archivePrefix = {arXiv},
       eprint = {1507.08285},
 primaryClass = {astro-ph.EP},
       adsurl = {https://ui.adsabs.harvard.edu/abs/2015PASP..127.1161K}
}

@ARTICLE{prose_2022,
       author = {{Garcia}, Lionel J. and {Timmermans}, Mathilde and {Pozuelos}, Francisco J. and {Ducrot}, Elsa and {Gillon}, Micha{\"e}l and {Delrez}, Laetitia and {Wells}, Robert D. and {Jehin}, Emmanu{\"e}l},
        title = "{PROSE: a PYTHON framework for modular astronomical images processing}",
      journal = {\mnras},
         year = 2022,
        month = feb,
       volume = {509},
       number = {4},
        pages = {4817-4828},
          doi = {10.1093/mnras/stab3113},
archivePrefix = {arXiv},
       eprint = {2111.02814},
 primaryClass = {astro-ph.IM},
       adsurl = {https://ui.adsabs.harvard.edu/abs/2022MNRAS.509.4817G}
}

@inproceedings{Hansen2008,
  author       = {Hansen, Marc P. and Malchow, Douglas S.},
  title        = {Overview of {SWIR} detectors, cameras, and applications},
  series       = {Society of Photo-Optical Instrumentation Engineers (SPIE) Conference Series},
  booktitle    = {Thermosense XXX},
  editor       = {Vladimir P. Vavilov and Douglas D. Burleigh},
  organization = {International Society for Optics and Photonics},
  publisher    = {SPIE},
  year         = {2008},
  volume       = {6939},
  doi          = {10.1117/12.777776},
  url          = {https://doi.org/10.1117/12.777776}
}

@ARTICLE{GJ1001,
       author = {{Golimowski}, David A. and {Henry}, Todd J. and {Krist}, John E. and {Dieterich}, Sergio and {Ford}, Holland C. and {Illingworth}, Garth D. and {Ardila}, David R. and {Clampin}, Mark and {Franz}, Otto G. and {Wasserman}, Lawrence H. and {Benedict}, G. Fritz and {McArthur}, Barbara E. and {Nelan}, Edmund P.},
        title = "{The Solar Neighborhood. IX. Hubble Space Telescope Detections of Companions to Five M and L Dwarfs within 10 Parsecs of the Sun}",
      journal = {\aj},
         year = 2004,
        month = oct,
       volume = {128},
       number = {4},
        pages = {1733-1747},
          doi = {10.1086/423911},
archivePrefix = {arXiv},
       eprint = {astro-ph/0406664},
 primaryClass = {astro-ph},
       adsurl = {https://ui.adsabs.harvard.edu/abs/2004AJ....128.1733G}
}

@ARTICLE{GJ1001temp,
       author = {{Dieterich}, Sergio B. and {Henry}, Todd J. and {Jao}, Wei-Chun and {Winters}, Jennifer G. and {Hosey}, Altonio D. and {Riedel}, Adric R. and {Subasavage}, John P.},
        title = "{The Solar Neighborhood. XXXII. The Hydrogen Burning Limit}",
      journal = {\aj},
         year = 2014,
        month = may,
       volume = {147},
       number = {5},
          eid = {94},
        pages = {94},
          doi = {10.1088/0004-6256/147/5/94},
archivePrefix = {arXiv},
       eprint = {1312.1736},
 primaryClass = {astro-ph.SR},
       adsurl = {https://ui.adsabs.harvard.edu/abs/2014AJ....147...94D}
}

@ARTICLE{Marana2025,
       author = {{Apergis}, Ioannis and {Bayliss}, Daniel and {Asimakoulas}, Leonidas and {Chote}, Paul and {McCormac}, James and {Mitchell}, Morgan A. and {Gill}, Sam and {Steen}, Philip G. and {Wheatley}, Peter J.},
        title = "{High-precision photometry with a scientific CMOS camera: I lab testing of the Marana camera}",
      journal = {RAS Techniques and Instruments},
         year = 2025,
        month = jan,
       volume = {4},
          eid = {rzaf049},
        pages = {rzaf049},
          doi = {10.1093/rasti/rzaf049},
archivePrefix = {arXiv},
       eprint = {2510.14484},
 primaryClass = {astro-ph.IM},
       adsurl = {https://ui.adsabs.harvard.edu/abs/2025RASTI...4af049A}
}

@BOOK{Janesick2001,
       author = {{Janesick}, James R.},
        title = "{Scientific charge-coupled devices}",
         year = 2001,
       adsurl = {https://ui.adsabs.harvard.edu/abs/2001sccd.book.....J}
}

@INPROCEEDINGS{Widenhorn2022,
       author = {{Widenhorn}, Ralf and {Blouke}, Morley M. and {Weber}, Alexander and {Rest}, Armin and {Bodegom}, Erik},
        title = "{Temperature dependence of dark current in a CCD}",
    booktitle = {Sensors and Camera Systems for Scientific, Industrial, and Digital Photography Applications III},
         year = 2002,
       editor = {{Blouke}, Morley M. and {Canosa}, John and {Sampat}, Nitin},
       series = {Society of Photo-Optical Instrumentation Engineers (SPIE) Conference Series},
       volume = {4669},
        month = apr,
        pages = {193-201},
          doi = {10.1117/12.463446},
       adsurl = {https://ui.adsabs.harvard.edu/abs/2002SPIE.4669..193W}
}

@INPROCEEDINGS{Sullivan2014,
       author = {{Sullivan}, Peter W. and {Croll}, Bryce and {Simcoe}, Robert A.},
        title = "{Near-infrared InGaAs detectors for background-limited imaging and photometry}",
    booktitle = {High Energy, Optical, and Infrared Detectors for Astronomy VI},
         year = 2014,
       editor = {{Holland}, Andrew D. and {Beletic}, James},
       series = {Society of Photo-Optical Instrumentation Engineers (SPIE) Conference Series},
       volume = {9154},
        month = jul,
          eid = {91541F},
        pages = {91541F},
          doi = {10.1117/12.2056425},
archivePrefix = {arXiv},
       eprint = {1406.2695},
 primaryClass = {astro-ph.IM},
       adsurl = {https://ui.adsabs.harvard.edu/abs/2014SPIE.9154E..1FS}
}

@ARTICLE{Zhang2023,
       author = {{Zhang}, Yihao and {Zhang}, Hongfei and {Liu}, Ziyi and {Zhu}, Jie and {Gu}, Jiayao and {Wang}, Xingbo and {Zhu}, Zeyu and {Tang}, Yujing and {Wang}, Jian},
        title = "{Glow and hot pixels removal using improved robust principal component analysis}",
      journal = {Journal of Electronic Imaging},
         year = 2023,
        month = jul,
       volume = {32},
          eid = {043012},
        pages = {043012},
          doi = {10.1117/1.JEI.32.4.043012},
       adsurl = {https://ui.adsabs.harvard.edu/abs/2023JEI....32d3012Z}
}

@article{Simcoe2019,
	author = {Simcoe, Robert A. and F{\H u}r{\'e}sz, G{\'a}bor and Sullivan, Peter W. and Hellickson, Tim and Malonis, Andrew and Kasliwal, Mansi and Shectman, Stephen A. and Kollmeier, Juna A. and Moore, Anna},
	doi = {10.3847/1538-3881/aae094},
	journal = {The Astronomical Journal},
	month = {jan},
	number = {2},
	pages = {46},
	publisher = {The American Astronomical Society},
	title = {Background-limited Imaging in the Near Infrared with Warm InGaAs Sensors: Applications for Time-domain Astronomy},
	url = {https://doi.org/10.3847/1538-3881/aae094},
	volume = {157},
	year = {2019}}

@article{Rogalski2002,
	author = {Antoni Rogalski},
	doi = {https://doi.org/10.1016/S1350-4495(02)00140-8},
	issn = {1350-4495},
	journal = {Infrared Physics & Technology},
	number = {3},
	pages = {187-210},
	title = {Infrared detectors: an overview},
	url = {https://www.sciencedirect.com/science/article/pii/S1350449502001408},
	volume = {43},
	year = {2002}}

@ARTICLE{Apergis2026,
       author = {{Apergis}, Ioannis and {Bayliss}, Daniel and {Chote}, Paul and {McCormac}, James and {Wheatley}, Peter J. and {Mitchell}, Morgan A. and {Fern{\'a}ndez Fern{\'a}ndez}, Jorge and {Gill}, Sam and {Bryant}, Edward M. and {Rodel}, Toby and {Asimakoulas}, Leonidas and {Anderson}, David R. and {Blake}, James A. and {Casewell}, Sarah L. and {Eeles-Nolle}, Fintan and {Hawthorn}, Faith and {Jenkins}, James S. and {Lendl}, Monika and {Lockley}, Isobel S. and {Moyano}, Maximiliano and {O'Brien}, Sean M. and {Saha}, Suman and {Smith}, Alexis M.~S. and {Steen}, Philip G. and {Vines}, Jose I. and {West}, Richard G. and {Zivave}, Tafadzwa},
        title = "{High-precision photometry with a scientific CMOS camera: II on-sky testing of the Marana camera at the NGTS facility}",
      journal = {RAS Techniques and Instruments},
         year = 2026,
        month = jan,
       volume = {5},
          eid = {rzag022},
        pages = {rzag022},
          doi = {10.1093/rasti/rzag022},
archivePrefix = {arXiv},
       eprint = {2603.16361},
 primaryClass = {astro-ph.IM},
       adsurl = {https://ui.adsabs.harvard.edu/abs/2026RASTI...5ag022A}
}

@ARTICLE{TOI,
       author = {{Guerrero}, Natalia M. and {Seager}, S. and {Huang}, Chelsea X. and {Vanderburg}, Andrew and {Garcia Soto}, Aylin and {Mireles}, Ismael and {Hesse}, Katharine and {Fong}, William and {Glidden}, Ana and {Shporer}, Avi and {Latham}, David W. and {Collins}, Karen A. and {Quinn}, Samuel N. and {Burt}, Jennifer and {Dragomir}, Diana and {Crossfield}, Ian and {Vanderspek}, Roland and {Fausnaugh}, Michael and {Burke}, Christopher J. and {Ricker}, George and {Daylan}, Tansu and {Essack}, Zahra and {G{\"u}nther}, Maximilian N. and {Osborn}, Hugh P. and {Pepper}, Joshua and {Rowden}, Pamela and {Sha}, Lizhou and {Villanueva}, Steven, Jr. and {Yahalomi}, Daniel A. and {Yu}, Liang and {Ballard}, Sarah and {Batalha}, Natalie M. and {Berardo}, David and {Chontos}, Ashley and {Dittmann}, Jason A. and {Esquerdo}, Gilbert A. and {Mikal-Evans}, Thomas and {Jayaraman}, Rahul and {Krishnamurthy}, Akshata and {Louie}, Dana R. and {Mehrle}, Nicholas and {Niraula}, Prajwal and {Rackham}, Benjamin V. and {Rodriguez}, Joseph E. and {Rowden}, Stephen J.~L. and {Sousa-Silva}, Clara and {Watanabe}, David and {Wong}, Ian and {Zhan}, Zhuchang and {Zivanovic}, Goran and {Christiansen}, Jessie L. and {Ciardi}, David R. and {Swain}, Melanie A. and {Lund}, Michael B. and {Mullally}, Susan E. and {Fleming}, Scott W. and {Rodriguez}, David R. and {Boyd}, Patricia T. and {Quintana}, Elisa V. and {Barclay}, Thomas and {Col{\'o}n}, Knicole D. and {Rinehart}, S.~A. and {Schlieder}, Joshua E. and {Clampin}, Mark and {Jenkins}, Jon M. and {Twicken}, Joseph D. and {Caldwell}, Douglas A. and {Coughlin}, Jeffrey L. and {Henze}, Chris and {Lissauer}, Jack J. and {Morris}, Robert L. and {Rose}, Mark E. and {Smith}, Jeffrey C. and {Tenenbaum}, Peter and {Ting}, Eric B. and {Wohler}, Bill and {Bakos}, G. {\'A}. and {Bean}, Jacob L. and {Berta-Thompson}, Zachory K. and {Bieryla}, Allyson and {Bouma}, Luke G. and {Buchhave}, Lars A. and {Butler}, Nathaniel and {Charbonneau}, David and {Doty}, John P. and {Ge}, Jian and {Holman}, Matthew J. and {Howard}, Andrew W. and {Kaltenegger}, Lisa and {Kane}, Stephen R. and {Kjeldsen}, Hans and {Kreidberg}, Laura and {Lin}, Douglas N.~C. and {Minsky}, Charlotte and {Narita}, Norio and {Paegert}, Martin and {P{\'a}l}, Andr{\'a}s and {Palle}, Enric and {Sasselov}, Dimitar D. and {Spencer}, Alton and {Sozzetti}, Alessandro and {Stassun}, Keivan G. and {Torres}, Guillermo and {Udry}, Stephane and {Winn}, Joshua N.},
        title = "{The TESS Objects of Interest Catalog from the TESS Prime Mission}",
      journal = {\apjs},
         year = 2021,
        month = jun,
       volume = {254},
       number = {2},
          eid = {39},
        pages = {39},
          doi = {10.3847/1538-4365/abefe1},
archivePrefix = {arXiv},
       eprint = {2103.12538},
primaryClass = {astro-ph.EP},
       adsurl = {https://ui.adsabs.harvard.edu/abs/2021ApJS..254...39G}
}

@inproceedings{pedersen2026duet,
  author    = {Pedersen, Peter P. and Baker, Clark E. and Dziewiecka, Karolina and Beck, Mathias and Queloz, Didier and Gillon, Micha\"{e}l and Triaud, Amaury H. M. J.},
  title     = {{DUET}: Design of a simultaneous {InGaAs}--{CCD} split-beam imager for ultra-cool dwarf exoplanet transit survey},
  booktitle = {Ground-based and Airborne Instrumentation for Astronomy XI},
  series    = {Proc. SPIE},
  volume    = {14149},
  year      = {2026},
  publisher = {SPIE},
  note      = {In press, paper 14149-438}
}



\newpage
\onecolumn
\appendix

\section{Target list}
\label{appendix}

\begin{table*}
\begin{minipage}{0.25\textwidth}
\centering
\begin{tabular}{|l| c c c c c|}
\toprule
Target Name & $T_\mathrm{eff}$ [K] & $zYJ$ mag & $J$ mag & $N_\mathrm{obs}$ & $t_\mathrm{exp}$\,[s] \\
\midrule
Sp0004-2058 & 2329 & 13.59 & 12.40 & 1 & 25 \\
Sp0004-4044 & 2041 & 14.45 & 13.11 & 7 & 31 \\
Sp0027-5401 & 2815 & 13.48 & 12.36 & 13 & 19 \\
Sp0102-3737 & 2648 & 12.47 & 11.13 & 2 & 6 \\
Sp0148-3024 & 3346 & 13.21 & 12.30 & 1 & 18 \\
Sp0158+1807 & 2760 & 12.82 & 11.97 & 3 & 14 \\
Sp0215-3040 & 2642 & 12.48 & 11.62 & 10 & 10 \\
Sp0218-0617 & 2724 & 13.73 & 12.85 & 1 & 26 \\
Sp0253-2336 & 2802 & 13.33 & 12.28 & 1 & 19 \\
Sp0253-7959 & 2850 & 12.12 & 11.34 & 27 & 8 \\
Sp0302-3950 & 3584 & 11.78 & 10.71 & 2 & 4 \\
Sp0314+2840 & 2850 & 11.88 & 10.99 & 6 & 6 \\
Sp0331-3042 & 2642 & 12.35 & 11.36 & 1 & 8 \\
Sp0334-2130 & 2828 & 12.99 & 11.92 & 8 & 13 \\
Sp0339-3525 & 2313 & 11.77 & 10.72 & 31 & 5 \\
Sp0435-1606 & 2500 & 11.21 & 10.41 & 1 & 3 \\
Sp0440-0530 & 2714 & 11.58 & 10.66 & 3 & 3 \\
Sp0544-2433 & 2605 & 13.56 & 12.53 & 6 & 22 \\
Sp0602-0915 & 2850 & 11.72 & 10.98 & 10 & 6 \\
Sp0704+0258 & 2720 & 13.40 & 12.53 & 5 & 19 \\
Sp0711-3824 & 2850 & 11.92 & 11.09 & 2 & 6 \\
Sp0740-1724 & 2803 & 10.96 & 10.15 & 2 & 3 \\
Sp0828-1309 & 2088 & 13.86 & 12.80 & 1 & 28 \\
Sp0833-5336 & 2784 & 12.91 & 12.02 & 10 & 13 \\
Sp0853-0329 & 2400 & 12.26 & 11.21 & 11 & 9 \\
Sp0931-1717 & 2825 & 11.99 & 11.07 & 1 & 6 \\
Sp1124-4213 & 2822 & 12.67 & 11.82 & 6 & 13 \\
Sp1145-2021 & 2850 & 10.40 & 11.73 & 16 & 12 \\
Sp1224-1238 & 2458 & 13.53 & 12.57 & 2 & 22 \\
Sp1255-5928 & 2850 & 11.99 & 11.15 & 12 & 7 \\
Sp1359-5636 & 2749 & 13.46 & 12.60 & 1 & 19 \\
Sp1421-1618 & 2698 & 13.66 & 12.76 & 4 & 23 \\
Sp1424-3514 & 2812 & 12.86 & 11.84 & 3 & 12 \\
Sp1504-2355 & 2765 & 12.90 & 12.01 & 5 & 14 \\
Sp1621-3711 & 2719 & 13.41 & 12.57 & 2 & 20 \\
Sp1718-4131 & 2831 & 11.37 & 10.61 & 1 & 4 \\
Sp1829-1439 & 2526 & 13.98 & 12.91 & 7 & 20 \\
Sp1836-3507 & 2666 & 13.48 & 12.14 & 1 & 16 \\
Sp1854-5704 & 2736 & 12.77 & 11.81 & 1 & 11 \\
\bottomrule
\end{tabular}
\end{minipage}
\hfill
\begin{minipage}{0.48\textwidth}
\centering
\begin{tabular}{|l | c c c c c|}
\toprule
Target Name & $T_\mathrm{eff}$ [K] & $zYJ$ mag & $J$ mag & $N_\mathrm{obs}$ & $t_\mathrm{exp}$\,[s] \\
\midrule
Sp1855-6914 & 2850 & 11.43 & 10.47 & 2 & 3 \\
Sp1921-2915 & 2744 & 12.31 & 11.45 & 11 & 8 \\
Sp1929+1101 & 2781 & 12.88 & 12.04 & 3 & 14 \\
Sp1945-2557 & 2605 & 13.15 & 12.35 & 1 & 21 \\
Sp2004-1416 & 2528 & 14.03 & 13.13 & 2 & 35 \\
Sp2038-5223 & 2808 & 13.94 & 12.95 & 1 & 32 \\
Sp2049-1716 & 2791 & 12.69 & 11.81 & 1 & 12 \\
Sp2151-4017 & 2650 & 12.27 & 11.45 & 6 & 8 \\
Sp2213-6342 & 2850 & 11.66 & 10.92 & 17 & 5 \\
Sp2226-7503 & 2793 & 13.34 & 12.35 & 7 & 18 \\
Sp2228-1325 & 2716 & 11.54 & 10.77 & 4 & 4 \\
Sp2259-5420 & 2712 & 13.01 & 12.25 & 1 & 16 \\
Sp2315-0627 & 2850 & 12.02 & 11.11 & 3 & 6 \\
Sp2331-2749 & 2573 & 12.55 & 11.65 & 3 & 11 \\
Sp2354-6312 & 2761 & 14.18 & 13.36 & 2 & 38 \\
TRAPPIST-1 & 2557 & 12.16 & 11.35 & 1 & 10 \\
TRAPPIST-1\,b & 2557 & 12.20 & 11.35 & 4 & 10 \\
TRAPPIST-1\,c & 2557 & 12.18 & 11.35 & 1 & 9 \\
TRAPPIST-1\,d & 2557 & 12.16 & 11.35 & 1 & 9 \\
TRAPPIST-1\,e & 2557 & 12.17 & 11.35 & 1 & 9 \\
TRAPPIST-1\,h & 2557 & 12.20 & 11.35 & 1 & 9 \\
\bottomrule
\end{tabular}
\end{minipage}
\caption{Summary of SPIRIT targets observed during the characterisation campaign. Effective temperature and $J$-band magnitude are taken from the SPECULOOS target list \citep[][]{Sebastian2021}; $zYJ$ magnitude is estimated from the photometry. TOI targets observed are excluded.}
\end{table*}

\newpage

\section{LP 944-20 time-series fit}
\label{appendixb}

\begin{figure}
\centering
\includegraphics[width=\linewidth, height=0.81\textheight, keepaspectratio]{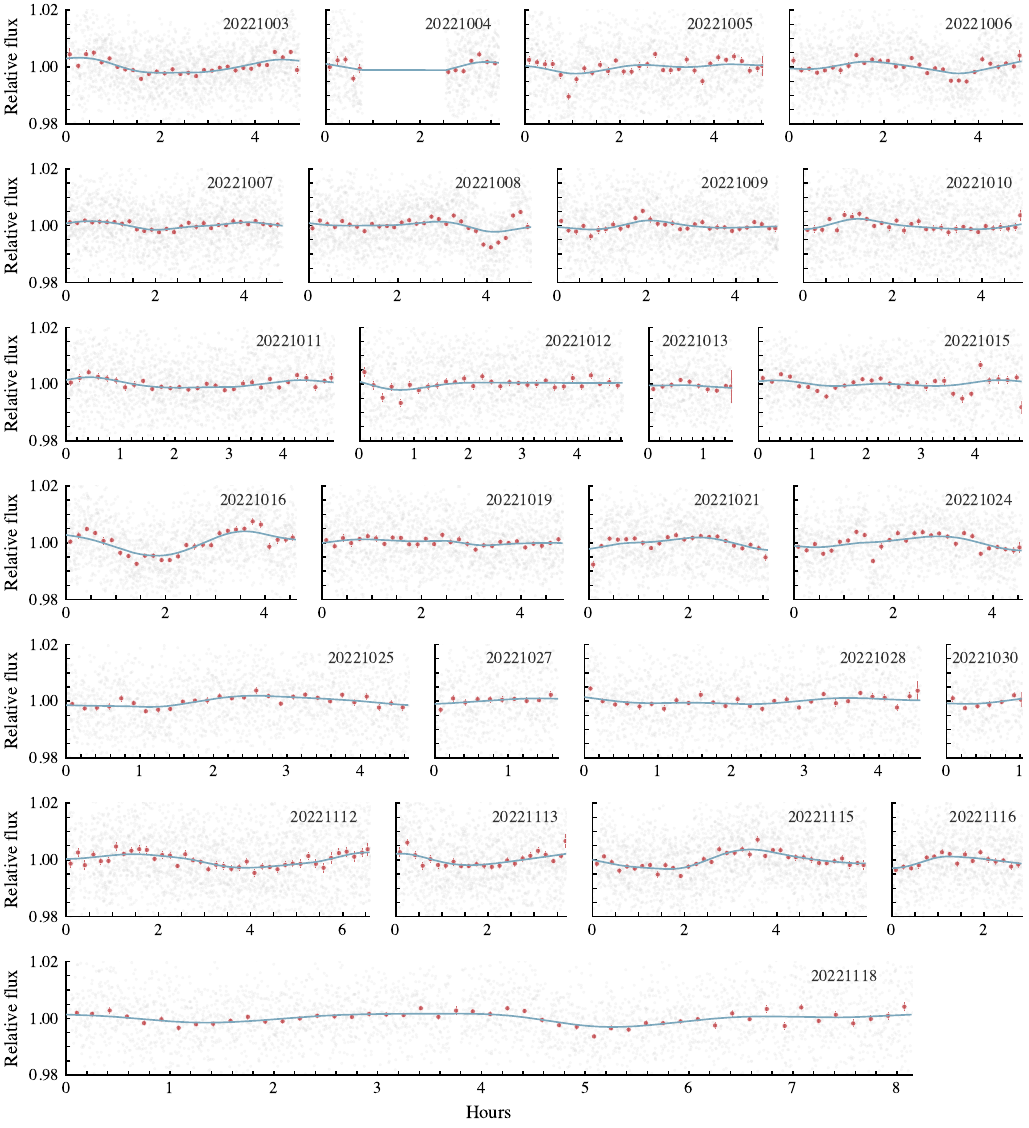}
    \caption{
    Individual SPIRIT $zYJ$-band light curves of LP 944-20 used in the rotation analysis. 
    Grey points show the unbinned photometry, red points show the data binned to 7-min cadence, and the blue curves show the corresponding best-fitting model including the detrending baseline. 
    Each panel represents a separate observing night, labelled by date in YYYYMMDD format, with time measured from the start of the observation.
    }
\end{figure}


\bsp	
\label{lastpage}
\end{document}